%% file: main.tex
\documentclass[sigconf]{acmart}

\AtBeginDocument{%
  }

\copyrightyear{2025}
\acmYear{2025}
\setcopyright{acmlicensed}\acmConference[CHI '25]{CHI Conference on Human Factors in Computing Systems}{April 26-May 1, 2025}{Yokohama, Japan}
\acmBooktitle{CHI Conference on Human Factors in Computing Systems (CHI '25), April 26-May 1, 2025, Yokohama, Japan}
\acmDOI{10.1145/3706598.3713435}
\acmISBN{979-8-4007-1394-1/25/04}

\usepackage{framed}
\usepackage{multirow}
\usepackage{float}
\usepackage{stfloats}

\newcommand{\inquote}[1]{``\textit{#1}''}
\begin{document}
\newcommand{\sys}{Toyteller}
\newcommand{\unfinalized}[1]{{#1}}
\newcommand{\john}[1]{\textbf{\textcolor{blue}{John: #1}}}
\newcommand{\maxk}[1]{\textbf{\textcolor{violet}{Max: #1}}}
\newcommand{\melissa}[1]{\textbf{\textcolor{brown}{Melissa: #1}}}

\title{\sys{}: AI-powered Visual Storytelling Through Toy-Playing with Character Symbols}

\author{John Joon Young Chung}
\affiliation{%
  \institution{Midjourney}
  \city{San Francisco}
  \state{CA}
  \country{USA}}
\email{jchung@midjourney.com}

\author{Melissa Roemmele}
\affiliation{%
  \institution{Midjourney}
  \city{San Francisco}
  \state{CA}
  \country{USA}}
\email{mroemmele@midjourney.com}

\author{Max Kreminski}
\affiliation{%
  \institution{Midjourney}
  \city{San Francisco}
  \state{CA}
  \country{USA}}
\email{mkreminski@midjourney.com}

\renewcommand{\shortauthors}{Chung et al.}

\begin{abstract}
\input{sections/00_abstract}

\end{abstract}


\begin{teaserfigure}
  \includegraphics[width=\textwidth]{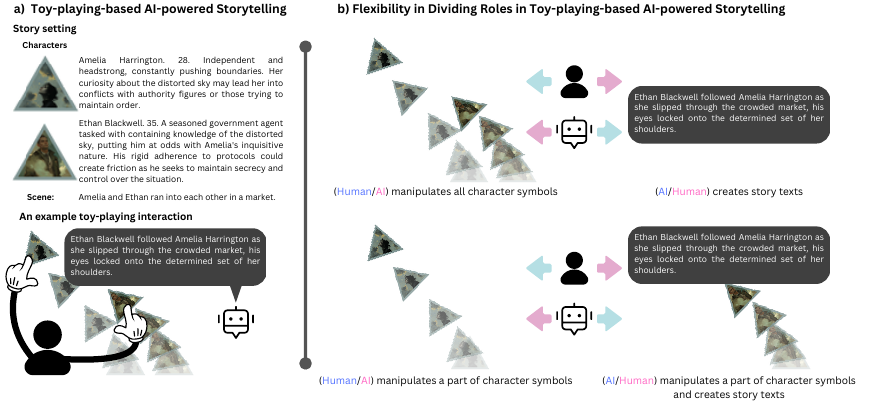}
  \caption{Interactions in \sys{}. a) \sys{} enables visual storytelling with AI through \textit{toy-playing} interaction with character symbols that accompany story texts. These character symbols can become both 1) a means to express the user's intention about the story unfolding and 2) a generation target that will be a part of the final visual story. For example, given a story setting defined as open-ended text, the user can physically manipulate symbols representing two story characters, and \sys{} can generate a story sentence that aligns with the user-provided character motions. b) \sys{} allows users flexibility in deciding which part of the output artifact (symbol motions, story texts) is created by the user or by the AI.}
  \label{fig:teaser}
\end{teaserfigure}


\maketitle
\input{sections/01_introduction}

\input{sections/02_related_work}
\input{sections/03_interaction}

\input{sections/04_technical}
\input{sections/05_technical_evaluation}
\input{sections/06_user_study}
\input{sections/07_design_space}
\input{sections/08_discussion}

\input{sections/09_conclusion}

\begin{acks}
We want to thank Midjourney for supporting this work.
We also want to thank Sarah Sterman for having a discussion about the project and Gawoo Kim for testing out \sys{}. 
\end{acks}

\bibliographystyle{ACM-Reference-Format}
\bibliography{main}

\appendix
\input{sections/10_appendix}
\end{document}

%% file: sections/00_abstract.tex
We introduce \sys{}, an AI-powered storytelling system where users generate a mix of story text and visuals by directly manipulating character symbols like they are \textit{toy-playing}. Anthropomorphized symbol motions can convey rich and nuanced social interactions; \sys{} leverages these motions (1) to let users steer story text generation and (2) as a visual output format that accompanies story text. We enabled motion-steered text generation and text-steered motion generation by mapping motions and text onto a shared semantic space so that large language models and motion generation models can use it as a translational layer. Technical evaluations showed that \sys{} outperforms a competitive baseline, GPT-4o. Our user study identified that toy-playing helps express intentions difficult to verbalize. However, only motions could not express all user intentions, suggesting combining it with other modalities like language. We discuss the design space of toy-playing interactions and implications for technical HCI research on human-AI interaction.

%% file: sections/01_introduction.tex
\section{Introduction}

Generative AI technologies, such as large language models (LLMs)~\cite{openai2024gpt4technicalreport, chowdhery2022palm}, have enabled many new forms of AI-supported storytelling~\cite{DSIIWA,chung2022talebrush, yuan2022wordcraft}. Many AI-powered storytelling applications, however, still rely heavily on natural language as the primary means of steering story generation---even though stories are often expressed through modalities beyond natural language~\cite{huang2016visual}. In human-human story co-creation, for example, children often casually create stories while they play with toys in their hands~\cite{benson1993structure, howes1992collaborative, talu2018symbolic}. Existing AI systems do not effectively support this kind of collaborative multimodal storytelling: even the most advanced AI models yet (1) exhibit a limited understanding of sequential multimodal inputs~\cite{chemburkar2024evaluating, zhang2024task}; (2) suffer from latency that limits interactive use~\cite{leviathan2023fast}; or (3) assume complex multimodal inputs (e.g., comic strips~\cite{jin2023generating}) that are difficult for users to manipulate casually. 

In this work, we introduce \sys{}, an AI-powered visual storytelling system that adopts the movement of character symbols as both an output modality and a steering input. Specifically, in \sys{}, the user can collaborate with AI to create the story of two characters along with their corresponding motions in symbols (Figure~\ref{fig:teaser}). The user manipulates the motion of one or two characters or writes down the story text, and AI fills in the rest of the content not filled in by the user (i.e., motion or text). \sys{} leverages the symbolic movements of characters in a manner inspired by Heider and Simmel’s experiments with the narrative interpretation of abstract shape animations~\cite{heider1944experimental}, as these anthropomorphized motions can express rich and nuanced semantics of social interactions through simple manipulation of symbolic objects. We achieve motion-to-text and text-to-motion generation by having a translational layer that maps motion and text inputs onto the action-specific subspace of a text embedding, which we use to condition large language models and motion generation models.

With a technical evaluation, we show that \sys{} is significantly better at enabling toy-playing interaction than competitive baselines backed by one of the most advanced large multimodal models, GPT-4o. Overall, \sys{} was more accurate in recognizing actions from motions, conditioning story text generation on motion inputs, and generating motions to match text. Moreover, across both text and motion generation, \sys{} did not hurt other qualities of generated artifacts but even improved some of them, such as the novelty and interestingness of text or the realism of motions. 
Lastly, \sys{} was faster than GPT-4o (e.g., up to $\times7.9$ and $\times557.4$ speed up for text generation and motion generation, respectively), enabling fluent real-time interactions. 

A user study of \sys{} found that toy-playing interaction fell into a different (and complementary) role to one of the most widely adopted AI steering approaches, natural language prompting. 
Participants found gestural toy-playing interaction useful for expressing ideas difficult to describe in words. They also perceived this interaction to be vague---which would be adequate to express underdeveloped ideas. However, due to its ambiguity and focus on motions, they found that expressing specific intentions with toy-playing can be difficult. As natural language prompts can express a lot of detailed information outside of motion, participants perceived that these approaches could complement each other. 
For instance, participants combined brief natural language prompts with the toy-playing input, to clarify the interpretation of toy-playing while spending less effort than expressing their intent via natural language alone.
\sys{}'s flexibility in dividing creation initiative between the user and AI allowed diverse complementary usages of two input modalities,
implying that \sys{} could support a range of varying user needs. We also identified suggestions for improving toy-playing interaction and its possible applications. 

Based on relevant previous work, our experience of designing \sys{}, and the user study, we explore the design space of toy-playing interactions for AI-powered storytelling in five dimensions: spatial mapping, temporal mapping, scene complexity, initiative division, and form factor. We conclude with reflections on how \sys{} extends AI-powered storytelling and how the HCI community could drive interactional and technical innovations in human-AI interaction research in the era of large AI models, along with technical alternatives for future work. 

In summary, this work contributes:
\begin{itemize}
    \item A novel interaction of toy-playing for AI-powered storytelling, which leverages the motion of character symbols as both a means to steer story generation and a visual accompaniment to a textual story. 
    \item \sys{}, a toy-playing-based AI-powered storytelling tool that enables motion-to-text and text-to-motion generation through a translational action information layer. 
    \item A technical evaluation and a user study which show how \sys{} surpasses GPT-4o in enabling toy-playing interaction and how it is used. 
    \item A design space and a reflection on toy-playing-based AI-powered storytelling, along with discussions on technical contributions and potential alternatives. 
\end{itemize}

%% file: sections/02_related_work.tex
\section{Related Work}

\begin{figure}[t]
  \includegraphics[width=0.3\textwidth]{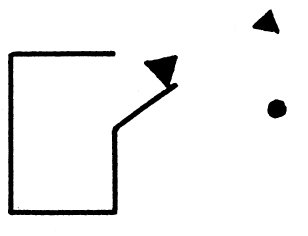}
  \caption{Heider and Simmel's experiment.}
  \label{fig:heider-simmel}
\end{figure}

\subsection{Toy-Playing for Storytelling}
Previous research has shown that humans can perceive actions or social interactions of characters in the motions of geometric symbols~\cite{gao2010wolfpack, gao2009psychophysics, heider1944experimental, barrett2005accurate}. 
In one especially prominent example, Heider and Simmel (Figure~\ref{fig:heider-simmel}) found that people readily anthropomorphize the motions of triangles and a circle, considering these shapes as characters and their motions as social interactions happening between these characters~\cite{heider1944experimental}. 
Abell et al. even leveraged such symbol motions to elicit mental states of observers in autism research~\cite{abell2000dotriangles}.
The existence of this recognition capability in people led researchers to investigate how computers might recognize similar information from symbol motions. Some of their efforts involved the collection of motion data---whether through touch interfaces~\cite{gordon2014anauthoring, roemmele2014triangle}, on a table-top interface~\cite{young2008puppet}, by attaching sensors to everyday objects~\cite{crick2008inferring}, or as a benchmark~\cite{maslan2015onehundred}. Based on the resulting datasets, researchers then designed algorithms to recognize actions from symbolic motions~\cite{roemmele2016recognizing, crick2008inferring, young2008puppet, singh2016recognizing, gordon2016commonsense}.

While AI has advanced rapidly since the first work in this direction, recent work found that advanced large vision-language models (GPT-4V~\cite{openai2024gpt4technicalreport} and Llava-1.5~\cite{liu2024improvedbaselinesvisualinstruction}) still have limitations in understanding actions from images of symbol motions~\cite{chemburkar2024evaluating}. 
Furthermore, previous work has largely aimed to classify motions strictly into one of several fixed, mutually exclusive action categories, thereby discarding many of the nuances (such as varying intensity of actions, or ambiguity in the action's intended outcome) that motion trajectories can convey. 
We introduce a storytelling system that leverages symbol motions for toy-playing interactions while overcoming technical limitations in previous work.

\subsection{AI-Powered Storytelling}
Building on progress in AI research, researchers and practitioners have introduced many tools to support storytelling experiences. 
Some of these tools support the definition of high-level story structure, for instance via tropes~\cite{chou2023talestream} or high-level plot events~\cite{kreminski2022loose}. As natural language generation capabilities improved, tools that directly suggest low-level story text started to emerge~\cite{roemmele2018automated, clark2018creative}, and a wide variety of similar text-focused tools proliferated with the advent of large language models~\cite{lee2022coauthor, yuan2022wordcraft, chung2022talebrush, mirowski2023cowriting, calderwood2020novelists}. While most of these focused on crafting story content in a fixed form, a thread of work enabled malleable story media where AI adaptively generates content personalized to the user's input or contexts~\cite{park2023generativeagents, aidungeon2_2019, kim2024authors}.
Researchers have also introduced tools to support story worldbuilding, generating elements (e.g., characters, factions, props) that consist of the world~\cite{qin2024charactermeet, chung2024patchview, kreminski2024intent}.
With AI's advancing capabilities to generate images, researchers explored approaches for AI-augmented visual storytelling~\cite{huang2016visual}. There have been diverse efforts, from comic strip creation~\cite{gong2023interactive} to creating the visual aspect of the story world~\cite{dang2023worldsmith}, generating images that can accompany story narrations~\cite{tan2024audioxtend}, and visual story creation for specific use cases, such as dream narratives~\cite{wan2024metamorpheus} and family expressive art therapy~\cite{liu2024whenhefeelscold}. Researchers also studied practicing writer's perspectives on using AI technologies~\cite{gero2023social, biermann2022from, kim2024authors, chakrabarty2024creativity} and conducted a literature survey on existing tools~\cite{DSIIWA}. We extend AI-powered storytelling by introducing character symbol motion as a modality that can be a part of visual stories while being a means to steer story generation.

\subsection{Multimodal AI Models}

Multimodal AI models have permitted the extension of storytelling support tools to forms outside of text. Early work in aligning multiple modalities includes approaches like extending an uni-modal model to multimodal usage~\cite{lu2019vilbert} and training models on a set of multimodal tasks~\cite{tan2019lxmert}, but one impactful approach was training a shared representation of multiple modalities with transformer~\cite{vaswani2017attention} encoders and contrastive loss~\cite{radford2021learningtransferable}. 
Contrastive loss, specifically, tries to minimize the distance between paired data in different modalities.
With this approach, starting with pairing text and images~\cite{radford2021learningtransferable}, researchers explored adding other modalities to text-image representation, including audio~\cite{guzhov2022audioclip}, 3D~\cite{zhang2022pointclip}, and even tactile data~\cite{yang2024bindingtouch}. Alternatively, researchers also trained multimodal representations from scratch with modalities other than text-image pairs, such as text-audio~\cite{elizalde2023clap}, text-video~\cite{xu2021videoclip} or even six modalities of images, text, audio, depth, thermal, and inertial data~\cite{girdhar2023imagebind}. Researchers used shared multimodal representations to enable generative tasks, for instance, by conditioning diffusion models for text-to-image~\cite{ramesh2022dalle2, saharia2022imagen}, text-to-audio~\cite{huang2023makeanaudio}, and text-to-3D~\cite{zhang2022pointclip} generation, or generating tactile data from other modalities and vice-versa~\cite{yang2024bindingtouch}.

However, using shared representations is not the only way to condition generation with inputs from other modalities. Other approaches include: 1) training a translational encoder that embeds one modality into embedding of another modality~\cite{liu2023llava}, 2) using an existing uni-modal representation as conditioning input for the generation of another modality~\cite{ho2022imagenvideo, kreuk2023audiogen}, and 3) training a unified transformer model that handles multiple modalities as discrete tokens~\cite{kondratyuk2024videopoet}. 
Learning from these, we enable motion-to-text and text-to-motion generation by training an encoder that maps motions to an existing text embedding space regarding character actions. 

%% file: sections/03_interaction.tex
\section{\sys{}: Interaction}

We first motivate toy-playing interaction in AI-powered visual storytelling and then explain \sys{}’s interface. 

\subsection{AI-Powered Storytelling via Toy-playing}
\label{sec:ai-powered-storytelling-via-toy-playing}

With toy-playing interaction, we leverage the human ability to perceive social interactions between characters, even from simple movements of symbolic shapes. The range of social interactions that symbolic motions can express is wide---for instance, Roemmele et al.~\cite{roemmele2016recognizing} categorized possible interactions between two characters into 31 classes. Moreover, even motions of the same category can express nuanced differences through varying dynamics, such as the magnitude or velocity of movement. For example, for a two-character motion representing a ``hit’’ action, the velocity of the hitting character might indicate how hard they hit the other.

In the context of AI-powered storytelling interactions, the movement of shapes can serve both as 1) a means to steer AI generation and 2) a target to create with AI. That is, as steering input, the user's manipulation of character symbols can condition the AI’s generation of story text. This steering input offers complementary benefits to other input forms (e.g., natural language prompts), allowing users to express nuanced intentions about story events via the relatively simple interaction of gestural object manipulation. Meanwhile, movements of character symbols can also serve as part of a story artifact, as a visual complement to the text. Hence, these movements can be a target of the AI generation. For instance, the AI can generate character motions from user-provided story text or as a reaction to the user's manipulation of other character symbols. As these character motions can serve as input and output, the user can flexibly decide which parts they would like to contribute to visual stories, and the AI can fill in the remaining parts based on whatever input the user supplies (Figure~\ref{fig:teaser}b). 

This \textit{toy-playing} interaction---humans and AI manipulating visual character symbols while telling accompanying stories---can have many usage scenarios. We introduce one application case, \sys{}, demonstrating AI-based toy-playing in a 2D screen interface. We discuss other possibilities and its design space in Section~\ref{sec:design_space}.

\begin{figure*}[t]
  \includegraphics[width=\textwidth]{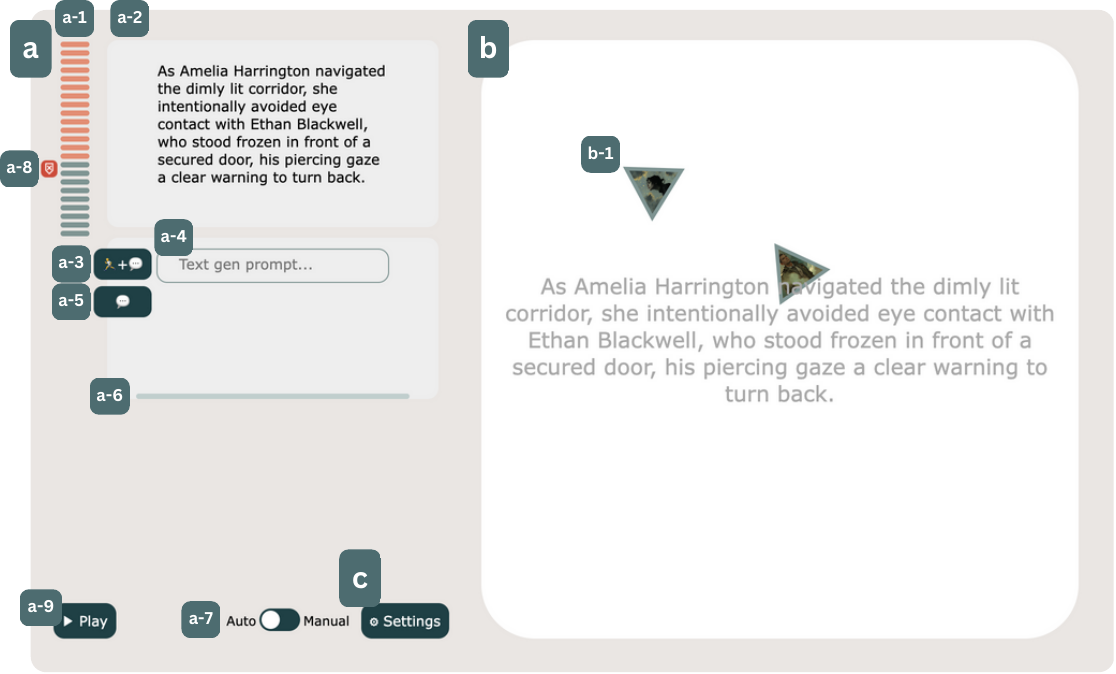}
  \caption{The \sys{} interface consists of a timeline module (a), playground (b), and a button to enter the setting page (c). The user can use the progress bar (a-1) to navigate recorded motion frames, which align with story textboxes (a-2). The user can manipulate character symbols (b-1) to record character motions and eventually initiate text generation. The user can ask \sys{} to generate both character motion and story text (a-3) while giving a high-level direction as a natural language prompt (a-4). The user can also initiate the generation of story text only (a-5) and adjust the size of the last textbox (a-6). By default, \sys{} automatically reacts to the user's input, but the user can turn this off (a-7). The user can delete frames and textboxes after a selected frame with a button (a-8). After recording frames and text, the user can play them back (a-9).}
  \label{fig:interface}
\end{figure*}

\subsection{Interface}

We designed \sys{} (Figure~\ref{fig:interface}) to demonstrate toy-playing-based AI-powered storytelling in a specific setting of interactions between two characters. While there can be more complex story settings with more characters and props, we consider dyadic settings as users can still express several stories with them. Moreover, \sys{} focuses on generating story prose text. \sys{} is designed for touchscreen and mouse-pointer interfaces, permitting the user to manipulate at most two character symbols simultaneously. 

\sys{} consists of three parts: 1) a setting page for configuring characters and story background; 2) a story timeline with a sequence of story text passages; and 3) a playground, 
which complements a text passage with user-manipulatable character symbols.

\begin{figure}[t]
  \includegraphics[width=0.478\textwidth]{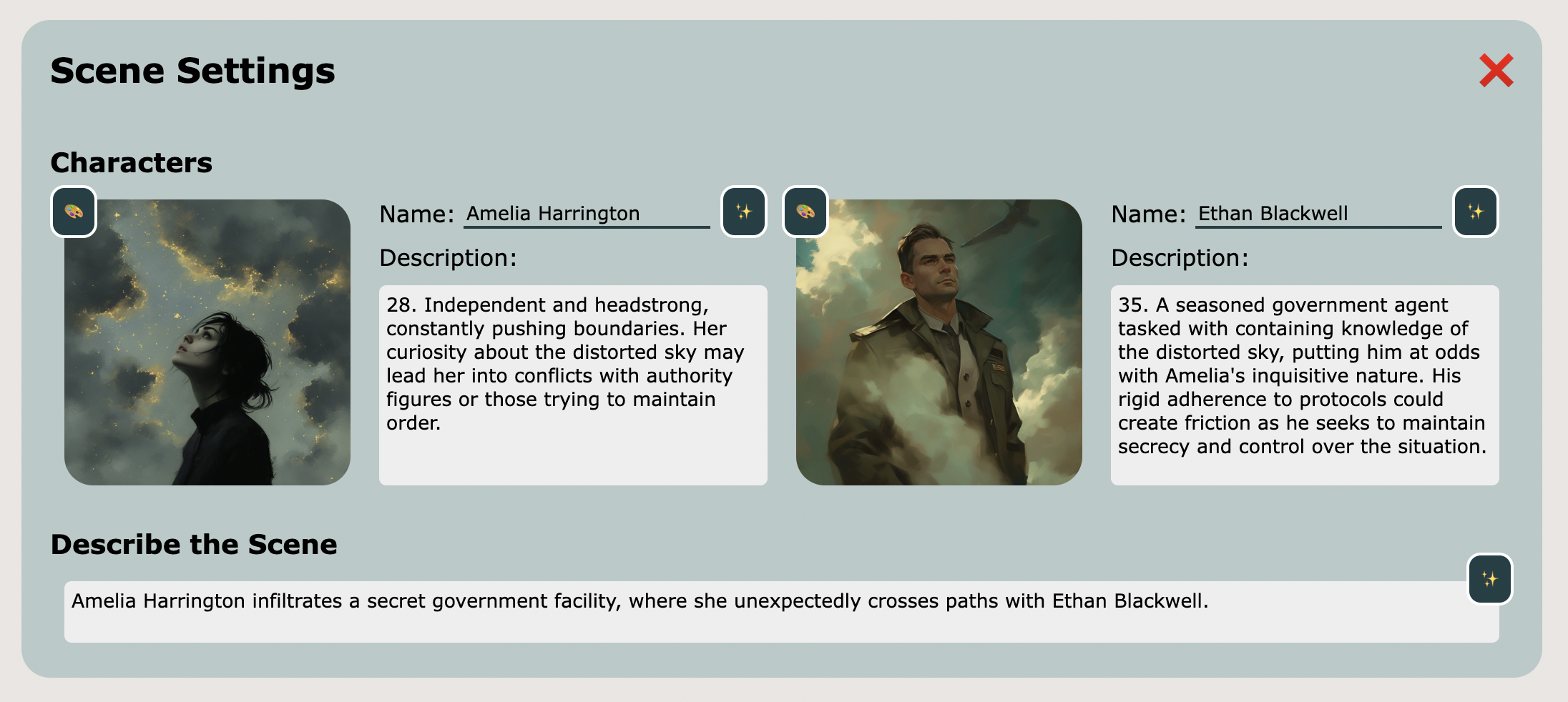}
  \caption{Interface for setting up the story scene. The paint and sparkle buttons allow users to use AI to generate setting-appropriate images and textual descriptions, respectively.}
  \label{fig:setting}
\end{figure}

\subsubsection{Setting}

Before unfolding the story with toy-playing interactions, the user needs to set the story scene by opening a setting page (with Figure~\ref{fig:interface}c). Specifically, the user can set 1) two characters’ names, 2) a brief textual description of each character, 3) a portrait of each character, and 4) a high-level description of the current scene (Figure~\ref{fig:setting}). \sys{} uses this information (except profile images) to guide the generation of story text, while images are overlaid on the character symbols in the playground to indicate which symbol corresponds to which character. 
While users can type in or upload these entries, \sys{} also provides the option of generating text and images with large language models and text-to-image models, with the sparkle and paint buttons, respectively.

\subsubsection{Story Timeline and Playground}

Once done with the story setting, the user can unfold the story with the story timeline (Figure~\ref{fig:interface}a) and the playground (Figure~\ref{fig:interface}b). The story timeline includes a progress bar that shows recorded motion frames (Figure~\ref{fig:interface}a-1) alongside story textboxes that correspond to certain ranges of recorded motion (Figure~\ref{fig:interface}a-2). The playground has two triangular character symbols, manipulatable by both the user and AI (Figure~\ref{fig:interface}b-1).

\paragraph{Motion $\rightarrow$ Text}
As mentioned in Section~\ref{sec:ai-powered-storytelling-via-toy-playing}, \sys{} allows flexible initiative division in creating visual stories. The first interaction approach involves the user and AI defining character motion first, then adding story sentences that align with this character motion. When contributing character motion, the user can decide to move one or two character symbols, and the AI will concurrently generate motion for uncontrolled character symbols. Alternatively, the user can invoke AI to generate motion for both characters (Figure~\ref{fig:interface}a-3). As the user or AI manipulates symbols, \sys{} records motion frames and accumulates them on the timeline. When the timeline reaches the end of the last textbox and the user is controlling one of the symbols, the size of the last textbox grows to align with the end of the timeline. As the user stops manipulating the symbols, \sys{} starts generating a story sentence that aligns with the provided motions. When AI generates both character motions, \sys{} starts generating the story sentence as the end of the timeline reaches the end of the last textbox. As toy-playing input cannot express all story ideas and can be vaguely interpreted, for the text generation, the user can prime the high-level direction with a natural language prompt (e.g., ``Write the ending of the story,'' Figure~\ref{fig:interface}a-4). The user can move to one of the recorded frames by clicking or scrubbing to the corresponding part in the timeline.

\begin{figure}[t]
  \includegraphics[width=0.478\textwidth]{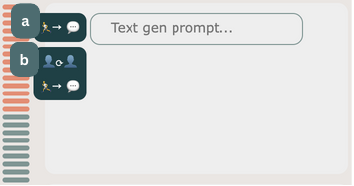}
  \caption{The user can manually initiate motion-to-text generation with a) and b). b) allows users to generate a sentence by swapping the ``active'' character of the event recognized by the system.}
  \label{fig:toyplaying-motion-to-text-interface}
\end{figure}

While the above interactions assume that \sys{} reacts to the user’s input automatically (e.g., as the user moves one character, AI moves the other character automatically), there might be cases where the user wants to decide the moment for AI generation. In such cases, the user can toggle the switch in Figure~\ref{fig:interface}a-7. As the user is in manual mode, if the user provides character motions first, they can manually generate story text with buttons in Figure~\ref{fig:toyplaying-motion-to-text-interface}. Note that the button in Figure~\ref{fig:toyplaying-motion-to-text-interface}b allows the users to swap who is active or passive in the event, as there are cases where AI fails to be correct in recognizing the active character in the event.

\begin{figure}[t]
  \includegraphics[width=0.478\textwidth]{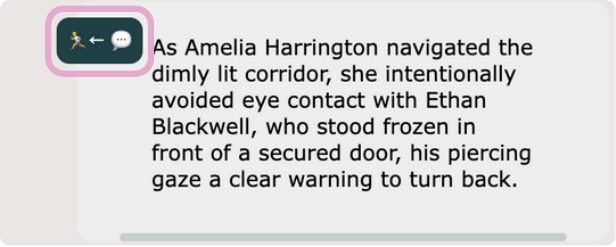}
  \caption{When text is provided before motions, with the highlighted button, the user can make AI generate motions.}
  \label{fig:toyplaying-text-to-motion-interface}
\end{figure}

\paragraph{Text $\rightarrow$ Motion}

The second way to interact with \sys{} involves the user or AI contributing passages of story text first, then creating character motions. Here, the user can write a story sentence by themselves or use AI to generate it (Figure~\ref{fig:interface}a-5). Again, when generating text with AI, the user can steer text generation via a natural language prompt. Then, the user can move some or all character symbols to align them with the provided sentence, and AI will generate the motion of uncontrolled characters. The user can also make the system generate all characters’ motions with the button in Figure~\ref{fig:toyplaying-text-to-motion-interface}. When the user is to adjust the textbox size to fit in fewer or more frames, they can use the handle in Figure~\ref{fig:interface}a-6.

\paragraph{Revision}

As a series of motions and story sentences accumulates, the user can revise them. For motions, the user can first move to the start of the frames they want to edit and manipulate the character symbols to override previously recorded motions. For sentences, the user can manually edit them in the text box. If they want to regenerate the sentence, they can delete the existing one first and use buttons in Figure~\ref{fig:toyplaying-motion-to-text-interface}. They can delete frames and sentences after a specific frame by moving to the frame in the timeline and then clicking the delete button (Figure~\ref{fig:interface}a-8). When done with editing the content, the user can play the recorded frames along with the aligned story sentences by clicking the \texttt{Play} button (Figure~\ref{fig:interface}a-9).

%% file: sections/04_technical.tex
\section{\sys{}: Technical Details}

We provide technical details of \sys{}, 1) the used dataset, 2) the system architecture to generate story text and symbol motion, and 3) implementation details of the interface and model training.  

\begin{table}[]
\caption{Action labels used in Roemmele et al.~\cite{roemmele2016recognizing}'s dataset}
\begin{tabular}{l}
\noalign{\global\arrayrulewidth=0.3mm}
\hline
\noalign{\global\arrayrulewidth=0.15mm}
\hspace{70pt}2-character action labels \\ 
\noalign{\global\arrayrulewidth=0.3mm}
\hline
\noalign{\global\arrayrulewidth=0.15mm}
\begin{tabular}[c]{@{}l@{}}accompany, approach, argue with, avoid, bother, capture, chase \\creep up on, encircle, escape, examine, fight, flirt with, follow \\herd, hit, huddle with, hug,  ignore, kiss, lead, leave, mimic \\play with, poke, pull, push, scratch, talk to, throw, tickle 
\end{tabular}
\\
\noalign{\global\arrayrulewidth=0.3mm}
\hline
\noalign{\global\arrayrulewidth=0.15mm}
\end{tabular}
\label{tab:dataset-action-terms}
\end{table}

\subsection{Dataset}

We used a part of Roemmele et al.’s ``Charades dataset’’~\cite{roemmele2016recognizing}. The dataset is composed of human-authored motions of one or two triangular character symbols. The current version of \sys{} uses only two-character motions, which can more readily express story events involving interaction between characters. Character motions in the dataset have an average length of 6.45 seconds, with a frame rate of 50 frames per second. Each frame contains considered character symbols' $x$-positions, $y$-positions, and rotations. Each instance's motion duration did not exceed 60 seconds.
Data instances are accompanied by action labels, with 31 ``base'' action terms used as categories (Table~\ref{tab:dataset-action-terms}). Each instance also labels one character as an ``active’’ character; this character is treated as the agent of the action, and the other character as the target of the action. We used 924 and 232 instances as training and test sets, respectively. We provide example data in the supplementary material.

\begin{figure}[t]
  \includegraphics[width=0.4\textwidth]{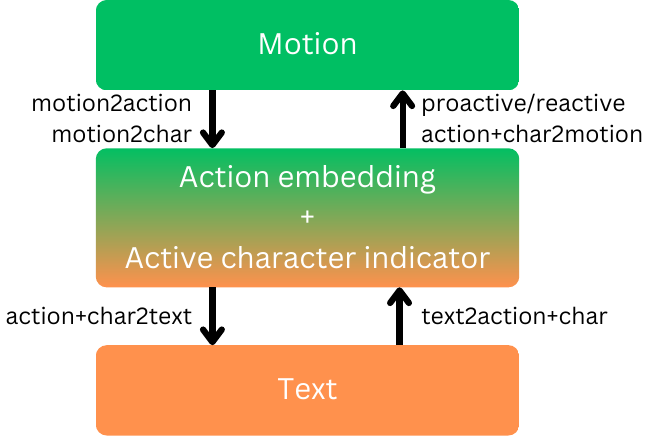}
  \caption{For flexible translation between motion and text, \sys{} leverages action information as a layer, which combines action embeddings and active character indicators.}
  \label{fig:toyplaying_technical_highlevel}
\end{figure}

\subsection{Model Architecture}

To enable flexible back-and-forth between motion and story text, we designed \sys{} to have an intermediate \textit{translational layer} (Figure~\ref{fig:toyplaying_technical_highlevel}). This translational layer serves as the semantic action vector to which motion and text can be projected. Moreover, this layer can serve as conditioning input for motion and text generation. In \sys{}, this translational layer consists of two pieces of information: 1) action embedding (\texttt{action}), the semantic vector that indicates \textit{which event is happening in between characters}, and 2) active character indicator (\texttt{char}), which is a boolean value of \textit{which character is the active agent of the action}. Note that these correspond to labels in the dataset we used, while we embedded action terms into the vector space with Sentence Transformer (a.k.a. SBERT)~\cite{reimers2019sentence-bert}. We trained models and designed pipelines to enable translations between this layer and motion/text:  \textit{motion $\rightarrow$ action}, \textit{action $\rightarrow$ motion}, \textit{text $\rightarrow$ action}, and \textit{action $\rightarrow$ text}.

\begin{figure}[t]
  \includegraphics[width=0.478\textwidth]{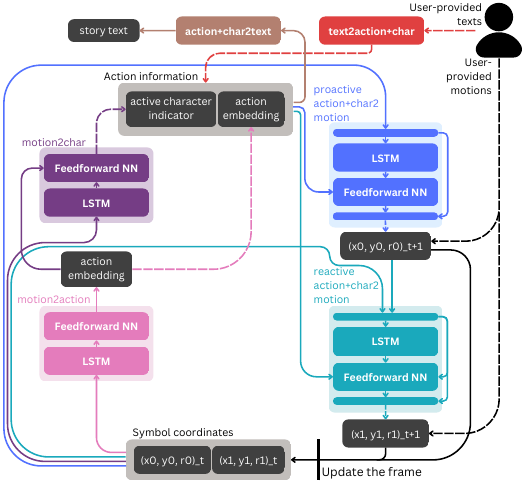}
  \caption{The architecture of \sys{} with information flow. \texttt{motion2action} and \texttt{motion2char} enable translation of motion to action information, while \texttt{proactive/reactive action+char2motion} generates motions out of action information. \sys{} turns action information into relevant story text with \texttt{action+char2text}. When the user provides a story sentence, \sys{} uses action information inferred from \texttt{text2action+char}.  
  Dashed lines indicate that the information can be provided either by \sys{} or the user.}
  \label{fig:toyplaying_technical_architecture}
\end{figure}

\subsubsection{Motions $\rightarrow$ Actions} 

To recognize action information, \sys{} uses two models, \textbf{\texttt{motion2action}} (Pink in Figure~\ref{fig:toyplaying_technical_architecture}) and \textbf{\texttt{motion2char}} (Purple in Figure~\ref{fig:toyplaying_technical_architecture}) models, which project motions into action embeddings and active character indicators, respectively. Both models combine an LSTM~\cite{hochreiter1997long} model and a feedforward neural network (NN).
LSTMs of both models use a series of coordinates for both characters ($x$-position, $y$-position, and rotation) as input to output internal state vectors. \texttt{motion2action}’s feedforward NN uses this vector as the input to infer the action embedding. Note that instead of using discrete action labels from the dataset, we trained \texttt{motion2action} model to infer the action’s vector in the SBERT’s semantic embedding space. This was due to 1) the ambiguity of interpreting actions from motions (e.g., motions to express hug and capture can look similar to each other) and 2) the continuity of embedding space that can potentially express nuances between actions (e.g., the intensity of an action based on how dynamic the motion is). \texttt{motion2char}’s feedforward NN used the concatenation of the LSTM's output vector and \texttt{motion2action}’s action embedding to infer the active character indicator. 

Note that, for processing motion frames, the current version of \sys{} uses LSTM over a widely adopted option, transformer architecture~\cite{vaswani2017attention}, due to the computation speed. Transformers slow down with longer sequences as they self-attend to all previous steps. On the other hand, LSTM maintains a hidden state to compress all previous steps and calculate the next step only with the current step and hidden state, which avoids any potential slow-down. We further discuss other technical options in Section~\ref{sec:technical_discussion}. 

\subsubsection{Actions $\rightarrow$ Motions}

As \sys{} infers the current action information, it can generate character symbol motions for the next frame. \sys{} first generates one character symbol ($sym_0$)’s coordinate with \textbf{\texttt{proactive action+char2motion}} model (Blue in Figure~\ref{fig:toyplaying_technical_architecture}). This model consists of a preprocessor, an LSTM, a feedforward NN, and a postprocessor. In the preprocessor, for the character symbol whose coordinates are not generated by this model ($sym_1$), the deltas of $x$ and $y$ positions are calculated, as the difference between the current frame ($t$) and the previous frame ($t-1$):
\begin{equation}
    ({dx}_{t}^{1}, {dy}_{t}^{1}) = ({x}_{t}^{1}, {y}_{t}^{1}) - ({x}_{t-1}^{1}, {y}_{t-1}^{1})
\end{equation}
The preprocessor also calculates the current distance between $sym_0$ and $sym_1$ as follows:
\begin{equation}
    ({xdist}_{t}^{1}, {ydist}_{t}^{1}) = ({x}_{t}^{1}, {y}_{t}^{1}) - ({x}_{t}^{0}, {y}_{t}^{0})
    \label{eq:dist1}
\end{equation}
Then, the LSTM of \texttt{proactive action+char2motion} uses (${dx}_{t}^{1}$, ${dy}_{t}^{1}$, ${xdist}_{t}^{1}$, ${ydist}_{t}^{1}$, ${r}_{t}^{0}$, ${r}_{t}^{1}$) as input to output an internal state vector, where $r$ denotes the rotation value of each character. 
Then, the feedforward NN uses the concatenation of the LSTM's output vector, action embedding, and active character indicator as the input, to infer (${dx}_{t+1}^{0}$, ${dy}_{t+1}^{0}$, ${r}_{t+1}^{0}$). The postprocessor turns this to the coordinate of $sym_0$ for the next frame, as below:
\begin{equation}
    ({x}_{t+1}^{0}, {y}_{t+1}^{0}, {r}_{t+1}^{0}) = ({x}_{t}^{0}, {y}_{t}^{0}, 0) + ({dx}_{t+1}^{0}, {dy}_{t+1}^{0}, {r}_{t+1}^{0})
\end{equation}
Note that when the user has provided the motion of this character, \sys{} replaces this generated value with the user-provided one.

With the $sym_0$'s motion provided either by \sys{} or the user, \sys{} then calculates the position of $sym_1$ with \textbf{\texttt{reactive action+char2motion}} model (Teal in Figure~\ref{fig:toyplaying_technical_architecture}). Similar to the proactive counterpart, this model also has a preprocessor, an LSTM, a feedforward NN, and a postprocessor. The input and output schemes are slightly different. The preprocessor first calculates $sym_0$'s position deltas and position difference between two characters:
\begin{equation}
    ({dx}_{t}^{0}, {dy}_{t}^{0}) = (x_t^0, y_t^0) - (x_{t-1}^0, y_{t-1}^0)
\end{equation}
\begin{equation}
    ({xdist}_{t}^{0}, {ydist}_{t}^{0}) = ({x}_{t}^{0}, {y}_{t}^{0}) - ({x}_{t}^{1}, {y}_{t}^{1})
    \label{eq:dist2}
\end{equation}
Then, the LSTM takes (${dx}_{t}^{0}$, ${dy}_{t}^{0}$, ${xdist}_{t}^{0}$, ${ydist}_{t}^{0}$, ${r}_{t}^{0}$, ${r}_{t}^{1}$) as the input to output an internal state vector. 
The feedforward NN takes the concatenation of the LSTM output, action embedding, active character indicator (swapped compared to the proactive one, as we consider another character), the deltas of $sym_0$'s $x$ and $y$ positions between the current and next frame (${dx}_{t+1}^0$, ${dy}_{t+1}^0$), and the rotation of $sym_0$ in the next frame (${r}_{t+1}^0$). The feedforward NN outputs ($dx_{t+1}^1$, $dy_{t+1}^1$, $r_{t+1}^1$), which is used to infer $sym_1$'s coordinate in the next frame:
\begin{equation}
    ({x}_{t+1}^{1}, {y}_{t+1}^{1}, {r}_{t+1}^{1}) = ({x}_{t}^{1}, {y}_{t}^{1}, 0) + ({dx}_{t+1}^{1}, {dy}_{t+1}^{1}, {r}_{t+1}^{1})
\end{equation}
Again, note that \sys{} uses this generated coordinate only when the user did not provide $sym_1$'s motions.

\begin{figure}[t]
  \includegraphics[width=0.478\textwidth]{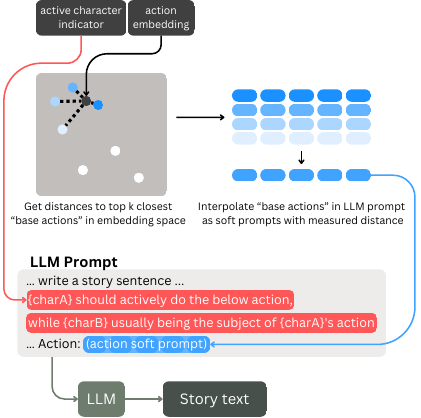}
  \caption{\texttt{action+char2text} pipeline. The pipeline first calculates the distance between the action embedding and base actions in the SBERT's embedding space. Then, it gets the soft prompt for the action embedding by weight-summing base action prompts in the LLM's input embedding space. \sys{} then uses this soft prompt along with an active character indicator to compose the prompt to the LLM.}
  \label{fig:toyplaying_technical_action2text}
\end{figure}

\subsubsection{Action $\rightarrow$ Text}

With \textbf{\texttt{action+char2text}} pipeline, \sys{} can use the recognized action information as the conditioning input for the story text generation (Brown in Figure~\ref{fig:toyplaying_technical_architecture}). To achieve this, \sys{} translates the action information into parts of the prompt to an LLM (Figure~\ref{fig:toyplaying_technical_action2text}). 
As \sys{} infers the action embedding as a continuous vector on the SBERT semantic space, instead of mapping this embedding to discrete input tokens for LLMs, \sys{} maps it to a soft prompt~\cite{lester2021power, li2021prefix, liu2023gptunderstands} in the LLM's input embedding space. Specifically, \sys{} computes a weighted sum of the base action terms (Table~\ref{tab:dataset-action-terms}) in the LLM's input embedding space, with weights coming from the cosine similarity between the base action terms and the inferred action embedding on the SBERT's vector space. When weight-summing, \sys{} considers the top $k$ most relevant base actions only, to assure that the resulting soft prompt only counts in highly relevant action terms. Moreover, as base action terms would have different LLM token lengths, we appended space (` ') tokens to some action terms so that all base actions can have the same token lengths when weight-summing them. 
\sys{} used this soft prompt in the LLM prompt as in the blue part of Figure~\ref{fig:toyplaying_technical_action2text}.
\sys{} also translated the active character indicator into the part of the prompt to indicate which character is active, as in the orange part of Figure~\ref{fig:toyplaying_technical_action2text}.
We provide the detailed prompt in Appendix~\ref{sec:prompt_actionchar2text}, and the LLM used the prompt to generate story sentences. This prompt also contains information such as character information, scene description, previous and following story sentences, and user-provided additional natural language prompts.

\subsubsection{Text $\rightarrow$ Action}

When the user provides story text, instead of using action information recognized from motions, through \textbf{\texttt{text2action+char}}, \sys{} infers action embeddings and active character indicators from provided text (Red in Figure~\ref{fig:toyplaying_technical_architecture}). \sys{} infers action embedding by the following procedure: 1) embed the user-given story text into the text embedding space with SBERT, 2) measure the distance between the embedded vector to the vectors for 31 base actions from the dataset, 3) filter top $k$ base actions that are closest to the embedded vector, and then 4) weight-sum top $k$ base action vectors with distance used as relative weights. We filtered only $k$ base actions to ensure the resulting action embedding considers only highly relevant actions. 
\sys{} recognizes the active character by prompting an LLM about which character is active in the story text regarding the inferred action embedding. 
To embed the recognized action embedding into the LLM prompt, with the same approach used in \texttt{action+char2text}, \sys{} turns the action embedding into the soft prompt in LLM's input embedding space. 
We provide the specific prompt that we used in Appendix~\ref{sec:prompt_text2char}.

\subsection{Implementation Details}
We implemented \sys{} as a web application built with HTML, CSS, and JavaScript. For the frontend and backend frameworks, we used React and Node.js, respectively. To handle AI-based operations, the backend of \sys{} communicates with a separate server built with Python and Flask. This server hosts \texttt{motion2action}, \texttt{motion2char}, \texttt{proactive action+char2motion}, and \texttt{reactive action+char2motion} models along with an LLM. 
Note that this server re-initiated the hidden states of these LSTM models for every sentence box (i.e., motions are processed at the sentence level) while caching previous hidden states.
This server also has \texttt{action+char2text} and \texttt{text2action+char} pipelines implemented. 
For the SBERT model for action embedding space, we used \texttt{all-MiniLM-L6-v2}, with its fair accuracy in calculating embedding similarity~\cite{sbert_eval}.
For the LLM, we used \texttt{\hyphenchar\font=`\-NousResearch/Meta-Llama-3-8B-Instruct}\footnote{\url{https://huggingface.co/NousResearch/Meta-Llama-3-8B-Instruct}}~\cite{llama3} through Huggingface Transformers library~\cite{wolf2020transformers}. 
Note that we needed to serve the LLM by ourselves to access the input embeddings and this model was small enough to fit on a single GPU for inference while having fair performance in a range of natural language tasks~\cite{llama3}.
\sys{} can use different LLMs if we can access and use their input embeddings. We served the models in the Flask server on a single \texttt{NVIDIA H100 SXM}.
We provide more details for different models and pipelines in the following sections. 
Note that we trained all models to handle symbol motions in 10 frames per second, by subsampling the dataset.
We did all model training on a single \texttt{NVIDIA H100 SXM}.

\subsubsection{\texttt{motion2action} and \texttt{motion2char}} We configured \texttt{motion\-2action}'s LSTM model with 4096 hidden features and eight recurrent layers. The feedforward NN had one hidden layer with 4096 features.\footnote{We used Pytorch (\url{https://pytorh.org/})'s \texttt{torch.nn.LSTM}, \texttt{torch.nn.Sequential}, \texttt{torch.nn.Linear}, and \texttt{torch.nn.ReLU} to implement these models.\label{ft:implementation}} We trained this model to output SBERT embeddings of the dataset's base action labels with mean absolute error loss. With the learning rate of $1e{-}5$ and batch size of eight, we trained the model for 50 epochs. Note that for all the models we trained, we picked the model checkpoint with the lowest test loss and used Adam as the optimizer. Moreover, for all models, we clipped the gradient norm at a maximum of five. We set \texttt{motion2char}'s LSTM model with 512 hidden features and four recurrent layers. The feedforward NN was with one hidden layer with 512 features. Using our dataset, we trained this model as a classification model with cross-entropy loss, a learning rate of $3e{-}5$, and a batch size of eight. We trained this model for 200 epochs.

\subsubsection{\texttt{proactive/reactive action+char2motion}} For both, LSTM models had 4096 hidden features with six recurrent layers. Feedforward NNs had one hidden layer with 4096 features.\footnotemark[2] We trained both models on our dataset with mean squared error loss. As these models generate a sequence of outputs, we trained them with a modified version of teacher forcing~\cite{williams1989alearning}. Specifically, for each frame, instead of using ground truth coordinates as input, we used the ``predicted coordinate'' which combines the previous frame's ground truth input and generated output. Specifically, when processing equations \ref{eq:dist1} and \ref{eq:dist2} to get training inputs, we followed the below equation, where overlines and tildes indicate the previous frame's ground truth and generated output, respectively:
\begin{equation}
    (xdist^i_t, ydist^i_t)=(\overline{x}^i_t, \overline{y}^i_t) - (\overline{x}^{|i-1|}_{t-1} + \tilde{dx}^{|i-1|}_{t}, \overline{y}^{|i-1|}_{t-1} + \tilde{dy}^{|i-1|}_{t})
\end{equation}
Through this, we intended to guide the model training while not biasing it too much to the training data.
We trained models for 200 epochs, with a batch size of 32. For learning rates, we used $1{-}e5$ for the proactive model and $1{-}e4$ for the reactive one.

\subsubsection{\texttt{action+char2text} and \texttt{text2action+char}}
For \texttt{action\-+char2text}, when filtering action terms most relevant to the current action embedding, we picked $k$ of four to ensure the soft prompt's relevance to the motion. For \texttt{text2action+char}, to filter action terms most relevant to the user-provided story text, we used $k$ of two, as text would be less ambiguous regarding the happening actions. With \texttt{\hyphenchar\font=`\-Meta-Llama-3-8B-Instruct}, we found that the maximum length of tokenized action terms was five. Hence, when weight-summing base action terms, for those action terms shorter than five tokens, we appended space tokens to make their length five. For temperatures used in the LLM, we used 0.7 for \texttt{action+char2text} and 0.0 for \texttt{text2action+char}.

%% file: sections/05_technical_evaluation.tex
\section{Technical Evaluation}
We technically evaluated \sys{} on its capability to 1) recognize action information from symbol motions, 2) generate story sentences conditioned on symbol motions, and 3) generate symbol motions from action information. By evaluating these, \textbf{we show our system outperforms a competitive baseline, GPT-4o (\texttt{gpt-4o-2024-05-13})~\cite{gpt4o}, in many aspects.}
We did not evaluate the pipeline to recognize action information from story text (\texttt{text2action+char}) due to the fair performance of the SBERT embedding (\texttt{all-MiniLM-L6-v2}) and the base language model (\texttt{Meta-Llama-3-8B-Instruct}). Specifically, the used SBERT embedding is fairly accurate in measuring embedding similarity~\cite{sbert_eval} and \texttt{Meta-Llama-3-8B-Instruct} is capable of doing a range of multiple-choice tasks~\cite{llama3}, which are specific tasks of \texttt{text2action+char}.

\subsection{Evaluating Motions $\rightarrow$ Actions}
We compared \texttt{motion2action} and \texttt{motion2char} to GPT-4o regarding recognizing action information from symbol motions. Specifically, using the test data from Roemmele et al.~\cite{roemmele2016recognizing}, we assessed these models' 1) ability to assign high rankings/weights to ``gold standard'' actions (\texttt{motion2action} vs. GPT-4o) and 2) accuracy in recognizing the active character (\texttt{motion2char} vs. GPT-4o). We also measured the models' latency, as fluent interactions would require low latency. To evaluate the models' capability to assign high weights to gold standard actions, we measured the ratio of the gold standard action weight to the weight of the top-ranked action. If the gold standard action is the top-ranked one, this metric would be one, while no weight assignment would turn into a zero. To calculate ranks and weights for \texttt{motion2action}, we used the cosine similarity between the inferred vector and the base action embeddings. For the baseline, we asked GPT-4o to provide weights in the order of ranks. Specifically, we used two types of motion inputs for GPT-4o: 1) a series of images rendered from coordinates (\texttt{GPT-4o-V}) and 2) coordinates in text (\texttt{GPT-4o-C}). Refer to Appendix~\ref{sec:gpt4ov-prompt-motion2action}, \ref{sec:gpt4oc-prompt-motion2action}, \ref{sec:gpt4ov-prompt-motion2char}, and \ref{sec:gpt4oc-prompt-motion2char} for specific prompts we used. For these prompts, we used 0.0 temperature. Note that, when evaluating active character recognition, we used the gold standard action in the dataset as the action input.  
For all evaluations of GPT-4o, we used zero-shot prompts, as adding examples would result in longer latencies. 

As ranks are ordinal values, we evaluated the differences in ranks through the Kruskal-Wallis test with Dunn's test as the post hoc test. For weights and latencies, as data did not have equal variance between conditions, we used Welch's ANOVA test to test the difference and then did Welch's t-tests for the post hoc test. Note that throughout this paper, we used Holm correction for the post hoc test. For the accuracy in recognizing active characters, we only measured the ratio of correctness without a statistical test.

\begin{figure*}[t]
  \includegraphics[width=\textwidth]{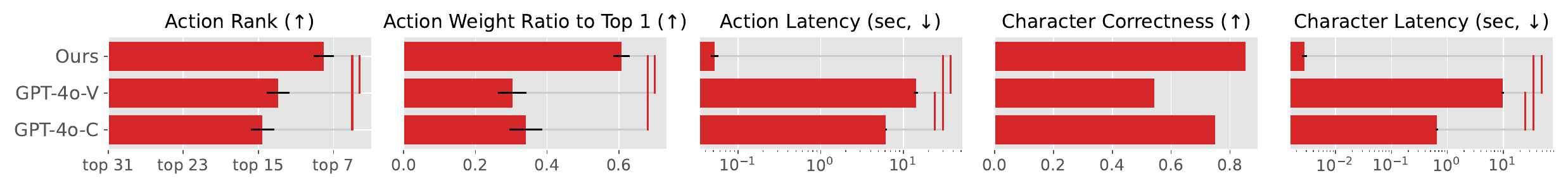}
  \caption{Comparing \sys{} (\texttt{Ours}) to \texttt{GPT-4o} approaches on inferring action information from motions. Specifically, we evaluated if \texttt{motion2action} places gold standard actions in higher ranks (Action Rank) with more weights (Action Weight Ratio to Top 1) in shorter times (Action Latency) compared to \texttt{GPT-4o} alternatives. We also compared \texttt{motion2char} to \texttt{GPT-4o} regarding accuracy in classifying active characters (Character Correctness) and latency (Character Latency). Throughout this paper, we marked significant differences with red bars and error bars indicate 95\% confidence intervals.}
  \label{fig:toyplaying_techeval_motion2action+char}
\end{figure*}

\subsubsection{Results}
\label{sec:motion2action+char_tech_eval}

Figure~\ref{fig:toyplaying_techeval_motion2action+char} shows the result. 
The three approaches were significantly different in how highly they ranked the gold standard action ($H(2)=65.69$, $p<0.001$). Specifically, \texttt{motion2action} ranked the gold standard actions significantly higher compared to \texttt{GPT-4o-V} and \texttt{GPT-4o-C} ($p<0.001$ for both). Weight assignment results also showed a significant difference between conditions ($F(2, 422.07)=111.99$, $p<0.001$), with \texttt{motion2action} assigning significantly higher relative weights to the gold standard action than other approaches ($p<0.001$ for both). 
Note that, however, \sys{} tended to assign weights in a more flat distribution than GPT-4o counterparts. 
Specifically, when we calculated Gini coefficient (a metric that indicates the concentration in a distribution) for weight assignments, \sys{}'s mean Gini coefficient (0.12) was significantly lower than GPT-4o counterparts (0.51 for \texttt{GPT-4o-V} and 0.60 for \texttt{GPT-4o-C}), indicating flatter distribution ($H(2)=475.46$, $p<0.001$ from Kruskal-Wallis test, and $p<0.001$ when comparing ours to GPT-4o conditions with Dunn's test). 
This result further motivates our approach of filtering top-k actions before interpolation, instead of incorporating irrelevant actions with high weights.
Moreover, there was a significant difference in the latency for recognizing \texttt{action}s ($F(2, 308.19)=2025.48$, $p<0.001$), with \texttt{motion2action} being drastically faster than GPT-4o approaches ($p<0.001$ for both). For recognizing active characters, \texttt{motion2char} was more accurate than GPT-4o approaches. In the latency of recognizing active characters, the difference between conditions was significant ($F(2, 617.38)=1166.13$, $p<0.001$), with \texttt{motion2char} being significantly faster than GPT-4o options ($p<0.001$ for both).

\subsection{Evaluating Motion $\rightarrow$ Text}
\label{sec:tech_eval_motion2text}
While the result in Section~\ref{sec:motion2action+char_tech_eval} shows \sys{} and GPT-4o's capabilities in recognizing action information from motions, it does not tell which approach is better at generating story text aligned with symbol motions. Hence, with human evaluators, we compared \sys{} to GPT-4o for motion-conditioned text generation. 

To generate story text with motion inputs, we first semi-randomly sampled 50 character motion instances from the test dataset. Here, to assure the diversity of motions, we made sure that the gold standard labels of sampled instances included all base actions. For character and setting inputs, we prompted GPT-4o to generate ten story settings with the prompt in Appendix~\ref{sec:generating-story-settings}. Note that we also provide generated settings in the appendix. We used generated characters and settings as inputs for story text generation. For \sys{}, we inferred action information with \texttt{motion2action} and \texttt{motion2char} models and then generated story text with \texttt{action+char2text}. In addition to the \sys{}'s original text generation pipeline (\texttt{Ours}), we also considered the condition that generates text only with the top-1 action without interpolating high-ranked actions (\texttt{Ours-top1}). 
With GPT-4o, we again examined both prompts that use a series of images (\texttt{GPT-4o-V}) and coordinate text (\texttt{GPT-4o-C}) as motion inputs, whose prompts are provided in Appendix~\ref{sec:gpt4ov-prompt-textgen} and \ref{sec:gpt4oc-prompt-textgen}, respectively. We used 0.7 as the temperature for all text generation.

We evaluated generated texts with three evaluators we hired from Upwork\footnote{\url{https://upwork.com/}}. Evaluators had earned at least \$7000 at the platform with 99\% task success rates, had experience in data annotation or creation, and were based in the US. Evaluators not only examined whether generated texts were well-aligned with the provided motion inputs but also other qualities of story texts. Specifically, we evaluated the creativeness in generated texts, which we phrased as novelty and interestingness in the questions, and the coherence, which we specified as coherence to the story setting and the grammatical correctness. 
We share specific evaluation questions in Appendix~\ref{sec:annotation_task}.
Evaluators rated these on 7-level Likert scales. For reliable evaluation, each evaluator rated all the generated instances, and for each instance, we aggregated all evaluators' ratings by taking means (resulting in 50 ratings for each condition and each metric).
We paid each evaluator \$60 for assessing 200 sentence-motion pairs, as each instance took around 1 minute (\$18/hr payment rate). We used Potato~\cite{pei2022potato} to deploy annotation tasks and randomized the orders of instances differently between evaluators. 

We also measured the latency to generate text and the diversity within all texts for each condition. Low latency would indicate more fluent interactions and high diversity would imply that the approach would likely prevent homogenization~\cite{anderson2024homogenization} in generated stories. For the diversity metric, we first embedded the story sentence with \texttt{all-MiniLM-L6-v2} and then calculated the dispersion of all instances based on minimum spanning tree (MST)~\cite{cox2021directed}.

For questions evaluators rated, as the data were in ordinal scale, we tested the significance with the Kruskal-Wallis test first and then conducted the post hoc test with Dunn's test. For the latency, as data had unequal variances, we did Welch's ANOVA test and then ran Welch's t-test as the post hoc test. As the diversity is calculated into one number, we could not conduct the statistical test for it. 

\begin{figure*}[t]
  \includegraphics[width=\textwidth]{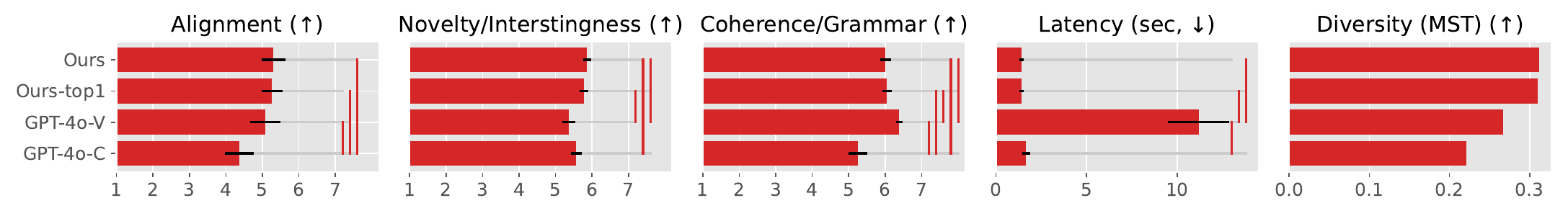}
  \caption{Comparing \sys{} (\texttt{Ours}) to the pipeline that does not interpolate actions (\texttt{Ours-top1}) and \texttt{GPT-4o} approaches on motion-conditioned story text generation. With human evaluators, we assessed motion-text alignment, novelty/interestingness, and coherence/grammaticality. We also evaluated latency and diversity in text generation.}
  \label{fig:toyplaying_techeval_motion2text}
\end{figure*}

\subsubsection{Results}

The results are in Figure~\ref{fig:toyplaying_techeval_motion2text}. Regarding the alignment of text to motion, the difference between conditions was significant ($H(3)=14.37$, $p{=}0.00244$). Our approach had the highest mean score in alignment, significantly better than \texttt{GPT-4o-C} ($p{=}0.00484$) but not significantly different compared to other approaches. With the novelty/interestingness scores, differences between conditions were significant ($H(3)=22.65$, $p<0.001$), with our approach having a significantly higher mean than \texttt{GPT-4o-V} and \texttt{GPT-4o-C} ($p<0.001$ and $p{=}0.0122$, respectively). Comparatively, \texttt{Ours-top1} was only significantly better than \texttt{GPT-4o-V} ($p{=}0.00508$). For coherence/grammaticality, conditions were significantly different ($H(3)=69.74$, $p<0.001$), with \texttt{GPT-4o-V} having a significantly highest mean than other approaches ($p<0.001$ for all). However, \texttt{Ours} and \texttt{Ours-top1} were significantly better than \texttt{GPT-4o-C} ($p<0.001$ for both). When we qualitatively inspected generated texts, \texttt{GPT-4o-V} generated mundane and short texts, which might have resulted in high coherence scores. With the latency, the differences between conditions were significant ($F(3, 102.80)=45.32$, $p<0.001$), with \texttt{GPT-4o-V} being significantly slower than other approaches ($p<0.001$ for all). With the diversity, \texttt{Ours} had the highest diversity, while the difference from \texttt{Ours-top1} was much smaller than those from GPT-4o alternatives.

\subsection{Evaluating Actions $\rightarrow$ Motions}

We evaluated \texttt{proactive} and \texttt{reactive action+char2motion} in terms of 1) alignment between conditioning actions and generated motions, 2) realism, whether the generated motions are natural without drastic jitters, and 3) latency. First, to generate motions, we semi-randomly sampled 31 instances from the test dataset, covering all base action categories. We used the gold standard action label as the conditioning input for motion generation. As each sampled instance has one character being active and the other being passive in the action, with each motion generation model, we generated two motion instances: 1) one generating the active character's motion while taking the data instance's passive character motion and 2) the other generating the passive character's motion along with the data instance's active character motion. Hence, with each model, we generated 62 motion instances. As the baselines, we used GPT-4o to simulate proactive and reactive motion generation, generating either passive or active character motions while taking the motion of counterpart characters from the test dataset. We list specific prompts we used in Appendix~\ref{sec:gpt4pro-prompt-motiongen} and~\ref{sec:gpt4re-prompt-motiongen}. Note that we ran these prompts to generate each frame's motion coordinate. We used 0.0 as the temperature. We only used the approach that considers motions as textual coordinates because generating motion images would likely take longer latency. As we configured the application to run in 10 frames per second, the motion generation latency should ideally be shorter than 0.1 seconds per frame. 

We requested the same human evaluators from Section~\ref{sec:tech_eval_motion2text} to assess action alignment and realism of generated motions in 7-level Likert scales. We share evaluation questions in Appendix~\ref{sec:annotation_task}. With 62 instances for each approach, evaluators rated 248 instances (proactive and reactive $\times$ our approach and GPT-4o). As each instance took around 50 seconds, we paid each evaluator \$70 (\$20.4/hr payment rate). We deployed the task with Potato~\cite{pei2022potato}. For each instance, we aggregated all evaluators' results by taking means.

As action-alignment and realism ratings are in ordinal scale, we used the Mann-Whitney U test for statistical testing. We tested latency with Welch's t-test as the data had unequal variance.

\begin{figure}[t]
  \includegraphics[width=0.478\textwidth]{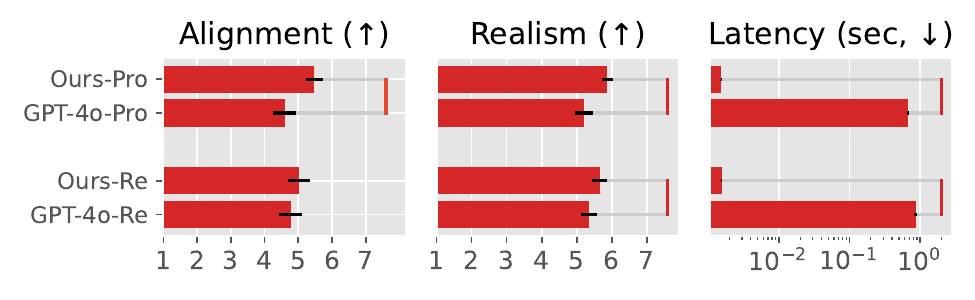}
  \caption{Comparing \sys{} (\texttt{Ours}) to \texttt{GPT-4o} on action-conditioned motion generation. \texttt{Pro} and \texttt{Re} indicate proactive and reactive motion generations, respectively. We evaluated these approaches with 1) alignment between actions and motions, 2) realism of the motion (e.g., the motion does not have drastic jitters), and 3) latency.}
  \label{fig:toyplaying_techeval_action2motion}
\end{figure}

\subsubsection{Results}

Figure~\ref{fig:toyplaying_techeval_action2motion} shows the results. When comparing proactive motion generation regarding alignment, \sys{} had significantly a higher mean score than GPT-4o ($U=2649.0$, $n_1=n_2=62$, $p<0.001$). With reactive approaches, \sys{} had a higher mean than GPT-4o, but the difference was not significant ($U=2120.0$, $n_1=n_2=62$, $p{=}0.321$). For realism, both comparisons on proactive and reactive approaches turned out to have significant differences, with our approaches having higher mean scores ($U=2680.0$, $n_1=n_2=62$, $p<0.001$ for proactive approaches, and $U=2340.5$, $n_1=n_2=62$, $p{=}0.0341$ for reactive ones). With latency, our approaches were significantly faster than GPT-4o alternatives ($t(61.00)=-67.22$, $p<0.001$ for proactive approaches, and $t(61.00)=-40.32$, $p<0.001$ for reactive ones). Note that our approaches took less than 0.1 seconds to generate a single frame while GPT-4o approaches took more than that. Considering that we designed \sys{} to work in 10 frames per second, only our approach would be reliable in generation speed.

%% file: sections/06_user_study.tex
\begin{figure}[t]
  \includegraphics[width=0.478\textwidth]{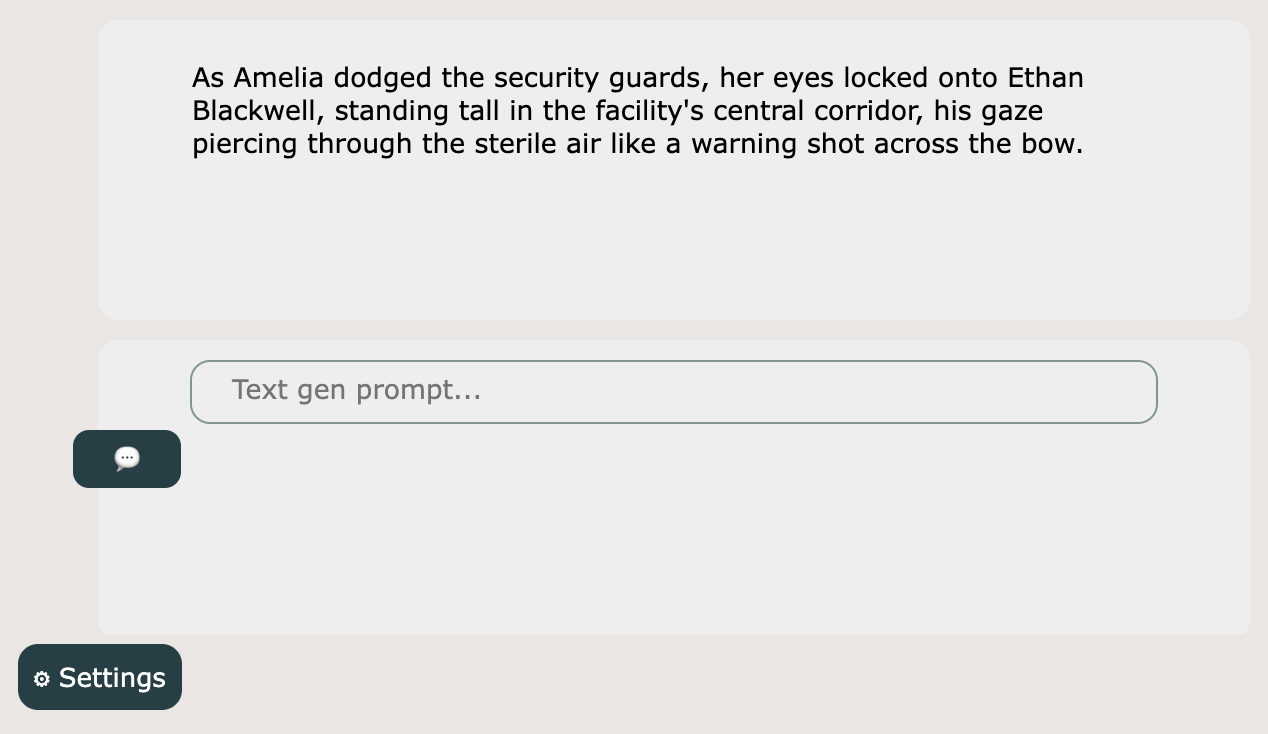}
  \caption{The baseline tool used in the user study.}
  \label{fig:toyplaying_baseline_tool}
\end{figure}

\section{User Study}

We conducted a user study to learn how \sys{}'s toy-playing interaction shapes the AI-powered storytelling experience. 
To better learn how toy-playing interaction is used, our study compared \sys{} to a baseline tool, which only uses natural language prompting as steering input to a generative AI model (Figure~\ref{fig:toyplaying_baseline_tool}). The baseline tool is a version of a tool that redacted toy-playing features from \sys{} and uses the same LLM as \sys{}.

\begin{table}[]
\caption{\unfinalized{Participant details}: story creation experience, story consumption frequency, and AI usage experience. The ``AI'' item indicates if they have ever used generative AI technologies. The ``AI for Create-Consume'' indicates whether they have experience using AI to create stories or enjoy AI-based story media content. ``Freq'' stands for ``frequently.'' }
\begin{tabular}{llllll}
\noalign{\global\arrayrulewidth=0.3mm}
\hline
\noalign{\global\arrayrulewidth=0.15mm}
 & Create & Consume      & AI & AI for Create-Consume \\ \noalign{\global\arrayrulewidth=0.3mm}
\hline
\noalign{\global\arrayrulewidth=0.15mm}
P1     & Novice & Frequently   & Y & N-Y (not freq) \\
P2     & Novice & Occasionally & Y & N-N \\
P3     & Novice & Occasionally & N & N-N \\
P4     & Novice & Occasionally & N & N-N  \\
P5     & Novice & Occasionally & Y & N-Y (not freq) \\
P6     & Hobbyist & Occasionally & Y & Y (freq)-Y (freq) \\
P7     & Novice & Frequently & N & N-N  \\
P8     & Expert & Frequently & Y & Y (freq)-Y (not freq)  \\
P9    & Novice & Frequently & Y & Y (freq)-Y (not freq) \\
P10     & Novice & Occasionally & Y & N-N \\
P11    & Hobbyist & Occasionally & Y & Y (not freq)-Y (not freq) \\
P12    & Expert & Occasionally & Y & Y (freq)-Y (not freq) \\

\noalign{\global\arrayrulewidth=0.3mm}
\hline
\noalign{\global\arrayrulewidth=0.15mm}
\label{tab:participant_details}

\end{tabular}
\end{table}

\subsection{Participants}

We recruited \unfinalized{12} participants (\unfinalized{seven women and five men, age 21-57, M=37.0, SD=10.8}) with Upwork\footnote{\url{https://upwork.com/}}. All participants were from the US. We did not limit the participants to those with story-creation experiences, as toy-playing interaction can be used to both create and consume stories (e.g., using toy-playing for consuming interactive narratives). However, before the study, we asked the participants about their story creation/consumption experiences and recruited only those with such experiences.
We also asked about their experiences with generative AI applications and whether they have used them for story creation/consumption. Table~\ref{tab:participant_details} details participants.

\subsection{Procedure}
We conducted a remote study, where participants joined Google Meet~\footnote{\url{https://meet.google.com}}. 
We requested participants to use tablets during the session since touch screens provide full control of Toyteller.
After welcoming the participants, we asked them if they were okay with recording the session. After starting the recording, we gave them the study overview and asked them to watch an instruction video on setting characters and a scene description. Then, participants created a story setting they would use later in the study. After that, we asked participants to use two versions of tools, the baseline and \sys{}, in randomized orders. For each condition, participants watched instruction videos and tried functions they learned from the video. 
With each tool, we asked participants to create a story of at least seven sentences with AI within 15 minutes. They could end the task early if they wanted to. Here, participants used characters and the scene description they had created earlier. During the usage, we recorded their tool usage screen and asked them to do a think-aloud to learn their thought processes. After each tool usage, to evaluate perceived creativity, participants completed survey questions on the creativity support index (CSI)~\cite{cherry2014quantifying} on a 0-to-10 scale. For the creativity support index, we excluded questions on collaboration as the tool was not for collaborative uses. After using both tools, we asked interview questions, which focused on 1) aspects they liked/disliked about each tool, 2) how they compare toy-playing with natural language prompting, and 3) how they can imagine using toy-playing interactions in practical story creation and consumption scenarios. Each participant session took no more than 70 minutes. We compensated each participant with \$35.

\subsection{Results}

To analyze CSI survey responses, we averaged scores of all relevant questions for each dimension and participant. Then, to compare two tools regarding each dimension, we conducted the Mann-Whitney U test. We used a non-parametric test as the original score data are in an ordinal scale.
We analyzed usage recordings and interview data through iterative coding with inductive analysis. 

\begin{figure*}[h]
  \includegraphics[width=\textwidth]{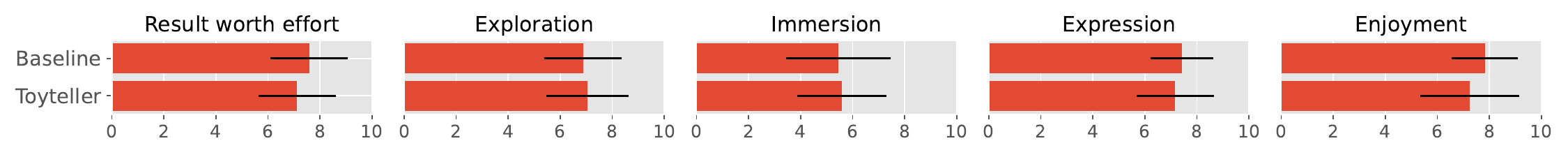}
  \caption{\unfinalized{CSI score results. No difference was statistically significant.}}
  \label{fig:toyplaying_csi}
\end{figure*}

\subsubsection{CSI scores}
As in Figure~\ref{fig:toyplaying_csi}, CSI scores between conditions did not have significant differences for all dimensions (\unfinalized{Mann-Whitney $U=81.5$, $p{=}0.602$ for result worth effort, $U=68.0$, $p{=}0.839$ for exploration, $U=70.5$, $p{=}0.954$ for immersion, $U=77.5$, $p{=}0.770$ for expression, and $U=74.5$, $p{=}0.907$ for enjoyment, with $n_1=n_2=12$}). We explain possible reasons for these results in the following sections with qualitative findings. 

\begin{figure}[t]
  \includegraphics[width=0.478\textwidth]{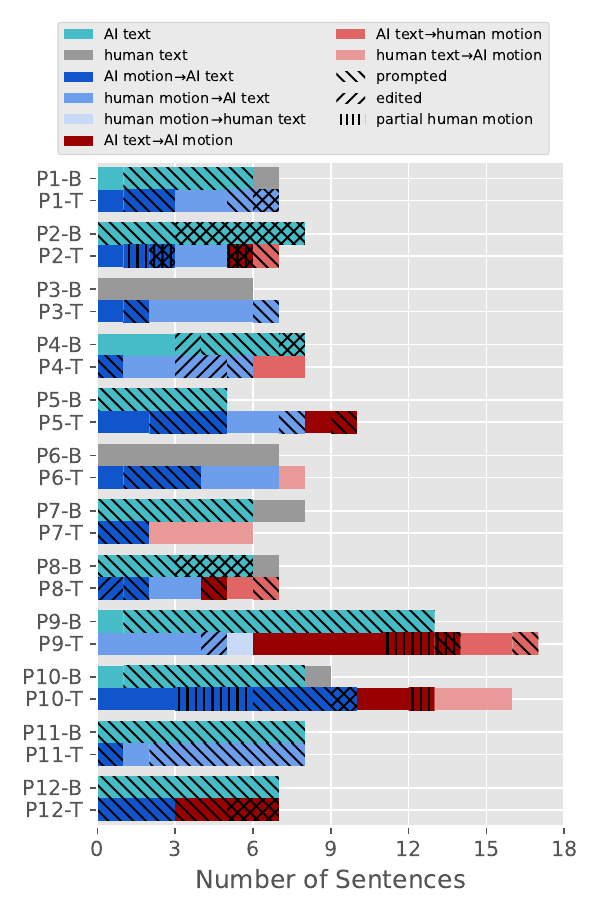}
  \caption{\unfinalized{The distribution of how participants created story sentences and motions.} `Prompted' means that the user input conditioning natural language prompts when generating text, and `edited' means that the user edited story text after AI generated them. `Partial human motion' indicates both human and machine contributed to creating motions. `B' and `T' denote `baseline' and `\sys{}', respectively.}
  \label{fig:toyplaying_usage_pattern}
\end{figure}

\begin{figure}[t]
  \includegraphics[width=0.478\textwidth]{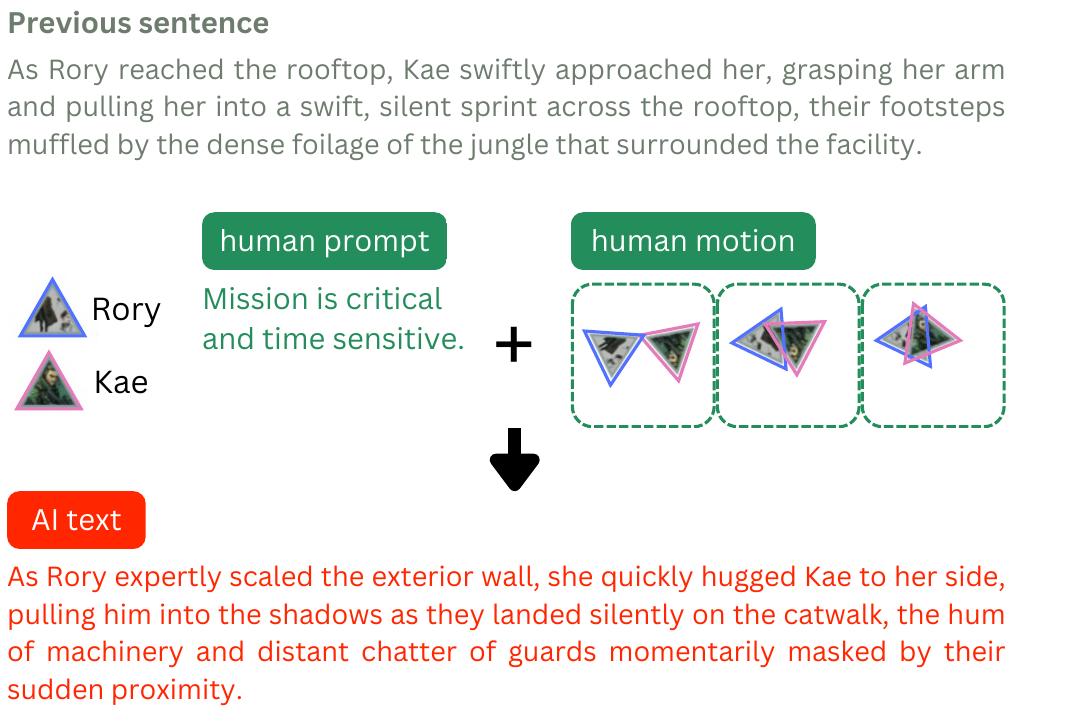}
  \caption{An example of mixing a natural language prompt with a motion input for text generation, from P5.}
  \label{fig:p5_prompt_motion_mix}
\end{figure}

\subsubsection{Toy-Playing and Natural Language Input Complement Each Other, as They are Vague and Specific, Respectively}
Participants thought of toy-playing interaction as \inquote{new adjectives that we might not think about before} (P2), which allows visual expression of ideas that are half-baked or difficult to describe verbally ($N=\unfinalized{8}$).
With such vague intentions conveyed via symbol motions, participants found the output of \sys{} to be surprising or unpredictable, which would be useful if they were looking for fresh inspirations. P2 mentioned: \inquote{I manipulated the motion ... I did not know what it was exactly going to generate. It was kind of hit or miss. Sometimes, toy-playing would generate something really good.}
However, with ambiguous and vague motion inputs, if the user had a narrow intention, they might disagree with the AI-generated text ($N=\unfinalized{2}$). 
Another limitation was that there could be cases where the user wanted to express ideas hard to describe with character motions ($N=\unfinalized{3}$).
On the other hand, participants thought that natural language inputs were adequate when they wanted to express more concrete and detailed ideas ($N=\unfinalized{10}$). P6 mentioned: \inquote{(If) you already have an idea of the next story scene, what you want to say, and how things are progressing along, you just type, type, and type.}
Considering different strengths, participants thought these approaches could complement each other ($N=\unfinalized{5}$). For example, the participant could use toy-playing to explore ideas and natural language prompts for concretization. 
Alternatively, some participants mentioned using toy-playing and natural language together, specifically when \inquote{you already have sort of a framework in your head, and you're trying to accent the text itself with the movement of the board} (P8).
We could observe such usage patterns, five participants mixing natural language prompts with motion inputs (light blue bars with hatches towards bottom right in Figure~\ref{fig:toyplaying_usage_pattern}), to either augment natural language prompts or narrow their intention regarding toy-playing.
Figure~\ref{fig:p5_prompt_motion_mix} shows an example, where P5 restricted the motion-conditioned text generation to a suspenseful one with an additional natural language prompt.
Furthermore, P6 thought the authors could have provided the overall story structures via prompts or text, and the audience could enjoy them as interactive narratives with toy-playing interactions.
Participants mentioned that users might vary in preferring these two approaches based on the specificity of their intentions and whether they are visual thinkers ($N=\unfinalized{4}$).
These different strengths might have been why CSI scores did not show significant differences: participants might have different goals and tastes and hence had different preferences over toy-playing and natural language prompts.

\begin{figure*}[t]
  \includegraphics[width=\textwidth]{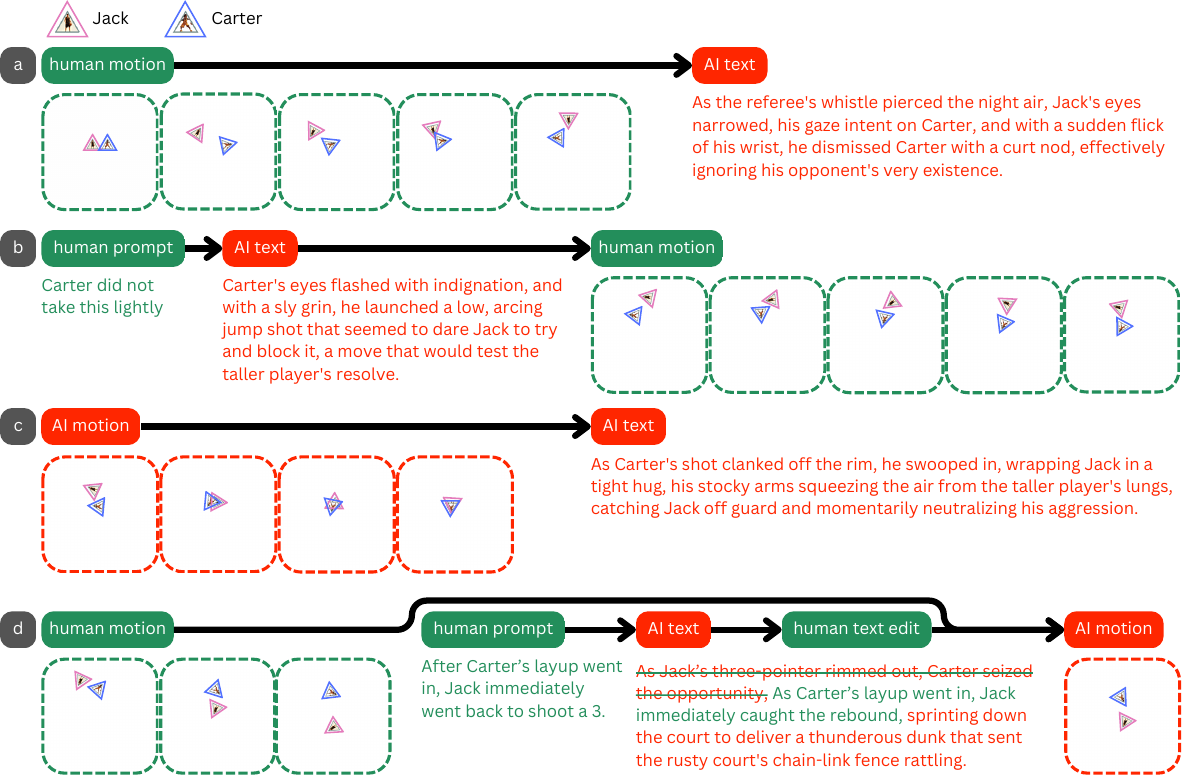}
  \caption{Examples of P2 co-creating story text and symbol motion with \sys{}.}
  \label{fig:toyplaying_p2_result}
\end{figure*}

\subsubsection{Various Usage Patterns}

Participants showed various ways to co-create story text and symbol motion with \sys{} (Figure~\ref{fig:toyplaying_usage_pattern}), which implies that \sys{} could support varying needs. 
We quantified how participants' ``creation approach'' distributions (Figure~\ref{fig:toyplaying_usage_pattern}) are different from each other when using \sys{}. Specifically, we measured an earth mover's distance equivalent for categorical distribution, the amount of the distribution weights that need to be moved to make one participant's distribution the same as another one's. The average of this distance between all pairs of participants was \unfinalized{0.61} (95\% CI: \unfinalized{[0.57 0.65]}). As weights for distributions were between zero and one, this value indicates that participants' creation approaches were highly different from each other.
We could observe that even a single participant adopted various creation approaches.
Figure~\ref{fig:toyplaying_p2_result} shows examples from P2. For instance, in Figure~\ref{fig:toyplaying_p2_result}a, P2 first created the motion of Jack avoiding Carter, and AI used it to generate the sentence where Jack is ignoring Carter. In Figure~\ref{fig:toyplaying_p2_result}b, P2 took the other way around, using AI to generate a story text with a conditioning prompt and then overlaying motion that would go along with the AI-generated text.
Figure~\ref{fig:toyplaying_p2_result}c is the case where P2 delegated all creation tasks to AI, with AI generating motion first and then text. In Figure~\ref{fig:toyplaying_p2_result}d, P2 made more contributions by themselves, from giving a prompt to condition text generation to editing the text and partially providing accompanying motions.
This variety of creation approaches could have been due to 1) providing steering interactions with different utilities (toy-playing and natural language prompting) and 2) the flexible design of \sys{} in dividing roles between the user and the system (Figure~\ref{fig:teaser}b).

\subsubsection{Improvement Suggestions for \sys{}} 
As \sys{} sporadically made errors in generating text or motion well aligned with user-provided input, further improving the technical pipeline was one suggestion ($N=\unfinalized{3}$). To participants, \sys{} also seemed to have its own bias in interpreting actions, not well adapting to stories with specialized domains, such as athletic stories ($N=\unfinalized{2}$). They thought making \sys{} more versatile to various settings would be necessary. Other suggestions were more about extending \sys{} features. Many participants mentioned improving the visual of toy-playing, as the current triangle symbols are limited in their expressivity ($N=\unfinalized{5}$). Participants suggested a broad range of new features, from adding more colors and shapes to generating reacting images or videos out of toy-playing interactions. Other suggestions included enabling more complex scenes with more characters or objects ($N=\unfinalized{3}$), limb motions ($N=\unfinalized{1}$), 3D arrangements ($N=\unfinalized{1}$), voice ($N=\unfinalized{1}$) or haptic ($N=\unfinalized{1}$) inputs, extended story settings ($N=\unfinalized{1}$), and generation of multiple options ($N=\unfinalized{2}$).

\subsubsection{Applications of AI-powered Toy-playing for Storytelling}
Many participants thought that toy-playing interaction would be useful in storytelling applications for children ($N=\unfinalized{5}$). For example, P10 mentioned that toy-playing interaction could even serve educational purposes for children: \inquote{For some people I know, they would find this a really interesting way to teach kids about writing and active voice and passive.} 
As toy-playing interaction focuses on the actions and motions of characters, participants considered it well-suited to movie or UX storyboarding ($N=\unfinalized{5}$). 
Interestingly, some participants mentioned that toy-playing could be useful for retrospection in the context of investigation or self-help, with the user describing their own experiences while doing the toy-playing and the AI inferring or concretizing aspects that the user may have overlooked ($N=\unfinalized{2}$). Participants also mentioned the possibility of using this technology to assist people with speech and language disabilities ($N=\unfinalized{2}$): \inquote{If you have somebody that is maybe non-verbal, interpreting their non-verbal, ``this is what they did,'' create an output, and the person of nonverbal can confirm or deny} (P8).

%% file: sections/07_design_space.tex
\begin{figure*}[t]
  \includegraphics[width=\textwidth]{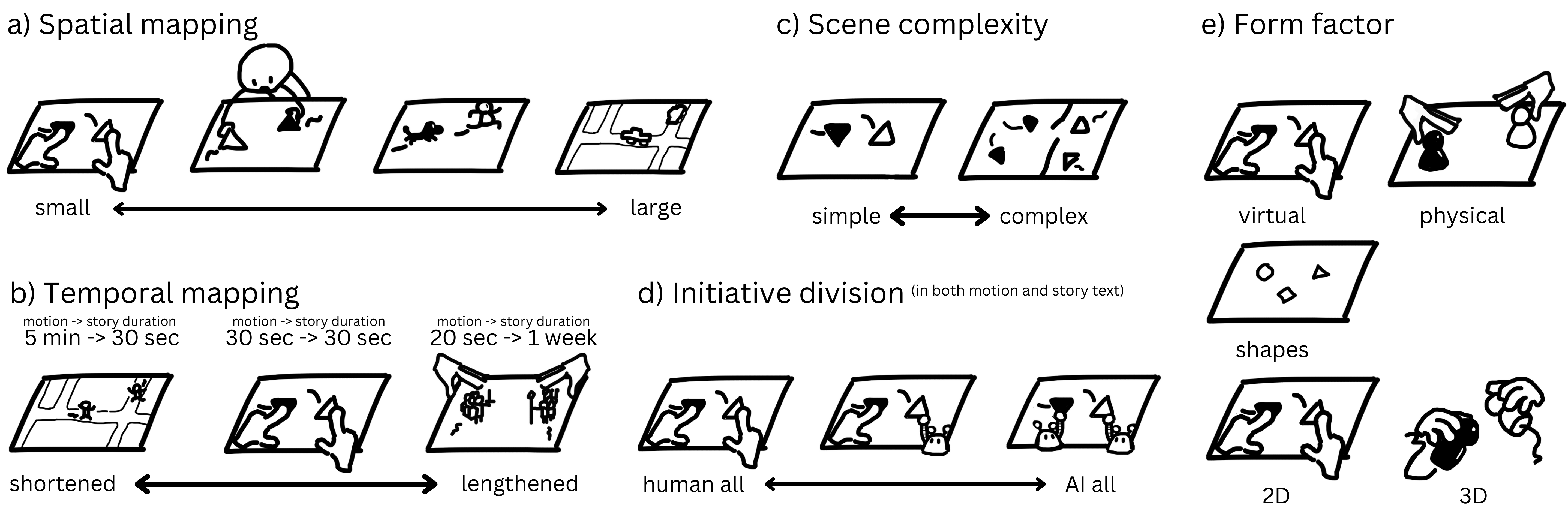}
  \caption{Design space for toy-playing interactions}
  \label{fig:toyplaying-designspace}
\end{figure*}

\section{Design Space of Toy-Playing Interaction}
\label{sec:design_space}

Based on previous work, our experience of designing \sys{}, and the user study, we discuss the design space of toy-playing interaction for AI-powered storytelling (Figure~\ref{fig:toyplaying-designspace}), how we can extend toy-playing beyond \sys{}.
We focus on five dimensions: a) spatial mapping, b) temporal mapping, c) scene complexity, d) initiative division, and e) form factor. Note that these dimensions are independently combinatorial if there is no technical restriction. Moreover, we explore how we can vary toy-playing designs rather than focusing on variations of textual or verbal stories themselves. The exploration of such combinations can be future work.

\subsection{Spatial Mapping}
While we implemented \sys{} on 2D screen devices (i.e., monitors and tablets), the scale of the interface where character symbols move can vary (Figure~\ref{fig:toyplaying-designspace}a). For instance, we can implement toy-playing on a tabletop interface~\cite{pedersen2011tangible, young2008puppet, kaimoto2022sketched}. Further increasing the scale of spatial mapping, we can match the movement of characters to the whole body motion of people, animals, everyday objects, or robots~\cite{crick2008inferring}. For example, we can have dogs with sensors in collars play on a field and let AI tell a story about what they are doing. To a further extreme, we can also imagine reading GPS signals from two people's cars and AI telling a reality-augmenting story based on their motions. As varying the spatial mapping, the dataset and models used for \sys{} might not apply directly as motion characteristics would depend on the size of the motion and who is performing the motion. Hence, technical adjustment might be necessary as we explore various spatial mappings.

\subsection{Temporal Mapping}
\sys{} assumed that the duration of motion almost matches to that in the story, but we can be flexible on the duration mapping (Figure~\ref{fig:toyplaying-designspace}b). Adjustment in temporal scale might be necessary sometimes due to large spatial scale. For example, assume that two people walk towards each other in a large area and take five minutes to physically contact each other. If we are to turn their motions into a story event of 30 seconds, we would need to map the motion duration to a shorter time in the story event. Alternatively, duration mapping can change due to the nature of story events. For example, if we use \sys{} for a war story, each event might take days or weeks in the story (e.g., relocation of troops), while the user might take only some seconds to express those events through motions. 
Some variations in temporal mapping might be achievable by adjusting story generation prompts (e.g., lengthening 20 seconds motions into an one-week story event), but others might require adjustment in models that handle motions (e.g., motions happening in longer duration than in a story due to larger spatial scale). 

\subsection{Scene Complexity}
\sys{} focused on dyadic interactions, but we can extend this to more complex scenes (Figure~\ref{fig:teaser}c). One direction is considering more than two characters. We can also consider other aspects, such as places and props in the world~\cite{hergenrader2018collaborative}. With more elements, the system would resemble role-playing games or AI-based simulations~\cite{park2023generativeagents}. With toy-playing as the core interaction, the system's capability to interpret inter-element motions in a short amount of time would be the core technical requirement to enable more complex scenes.

\subsection{Initiative Division}
\sys{} allowed flexible division in initiatives~\cite{DSIIWA} between AI and the user (Figure~\ref{fig:teaser}b), but flexibility might not always be required for toy-playing-based applications (Figure~\ref{fig:toyplaying-designspace}d). For instance, for toy-playing-based role-playing games, the system would allow the user to control only one character while AI would take the role of others as non-playable characters. Similarly, if we are to simulate a story world while considering the user as the audience outside of the world, we would need full automation. Note that we can configure different initiative divisions along the dimensions of different modalities. For example, in Figure~\ref{fig:teaser}a, the user moves characters while the AI narrates the story on behalf of the user. 

\subsection{Form Factors}

While \sys{} implemented toy-playing interactions in a 2D virtual screen, there can be various form factors for them (Figure~\ref{fig:toyplaying-designspace}e).
As implied in previous dimensions, toy-playing interactions can happen in physical worlds, with everyday objects or bodily motions. For instance, we can implement toy-playing interactions on a tabletop interface with physical robots. The shape of character symbols can also be one design aspect. When setting the shape of forms, it would be desirable to set the shape in a way that can provide affordances to express all motion information. For example, a circle cannot express rotation and hence might need to be avoided if the system designer wants to consider it in their applications. 
We can also consider manipulatable shapes, such as limbs that are controllable by either users or AI~\cite{jiang2024motiongpt}, which can further augment intra-toy motions.
Note that we can flexibly adjust the shape of virtual objects while physical ones would require more costs for such adjustments.
Moreover, while we implemented \sys{} for two-dimensional motions, we can imagine extending toy-playing to incorporate three-dimensional motions. With 3D motions, toy-playing would express a richer and more nuanced set of actions, such as jumping or attacking from below. For this, collecting 3D motion data and training models with such data would be necessary.

%% file: sections/08_discussion.tex
\section{Discussion}

\subsection{AI-Powered Storytelling with Toy-playing: Limitations and Extensions}

We found that toy-playing interaction supports storytelling differently than natural language prompting, in particular by allowing users to gesturally express ideas that are underdeveloped or difficult to verbalize~\cite{chung2022gestural}. 
Due to the vagueness of toy-playing, however, expressing all user intents only with toy-playing interactions can be difficult.
How can we best leverage these vague input approaches in the design of future AI-powered storytelling tools? We suggest that toy-playing interaction would be clearly appropriate in tools aimed purely at supporting ideation, as it allows users to express half-baked ideas through simple gestures. Meanwhile, in tools that support a more comprehensive story development process (from ideation to concretization), toy-playing interaction would be most effective for a specific subset of tasks. 
Hence, it would be important to let the users know when toy-playing interaction could be helpful. Giving tutorials or intelligently recommending the right features could be possible approaches. Using natural language and toy-playing together can be another approach, given that this leverages the complementary strengths of each modality. Toy-playing enables more nuanced expression of character motion compared with prompting, while prompting illuminates details outside of the motion dimension captured by toy-playing. 
To reduce the friction of switching between these input types, we can adopt a natural language modality that can be used concurrently with toy-playing, such as voice input. For instance, when augmenting the user's toy-playing with AI storytelling, the user can talk aloud during toy-playing (e.g., one character's line, such as ``I don't like you'') to disambiguate their intention.
Alternatively, the application could allow users to explore a limited story space with toy-playing by setting the overall narrative flow beforehand. For instance, content authors could script one or several possible plots beforehand, and the audience could explore those via toy-playing.
Future work might involve more evaluation of toy-playing interactions in these specific task settings, investigating whether they improve the user's creative experience compared to other interaction approaches.

We hope that this work will inspire researchers to introduce novel interactions in AI-powered storytelling. One strategy is bringing in human activities outside of AI-powered storytelling into the technical interaction realm~\cite{helander1997theroleofmetaphors, sharp2007interaction}. \sys{}, for instance, got inspiration from children imagining stories while playing with toys.

\sys{} could also be combined with other research efforts in the computational storytelling community. For instance, \sys{} could use story characters and settings created with AI-supported worldbuilding tools~\cite{qin2024charactermeet, chung2024patchview, kreminski2024intent}. Moreover, \sys{}'s quality of story generation could be improved by combining approaches like hierarchical generation~\cite{mirowski2023cowriting}, prompting with story plans from symbolic algorithms~\cite{wang2024guiding}, or suspenseful story unfolding~\cite{xie2024creatingsuspenseful}. At the same time, LLM-based story generation might incur homogenization~\cite{anderson2024homogenization}, which could potentially be mitigated through techniques for diversifying text generation~\cite{wang2024guiding, zhang2024forcingdiffuse}.

\subsection{Technical Contributions and Alternatives}
\label{sec:technical_discussion}

Large, pretrained, general-purpose AI models have made it substantially easier to efficiently prototype novel HCI systems. Prompt engineering in particular has been especially widely adopted, because it permits rapid investigation of diverse interactional ideas simply by overlaying human interfaces on prompt-engineered models. However, researchers have also criticized purely prompt-engineering approaches to prototyping and expressed concern about increasing overreliance on pretrained large models in HCI system-building~\cite{arawjo2024llm}. The most critical limitation would be that our interactional ideas would be bound to the capabilities of these models. 

Our research shows one way to build upon large general-purpose models to introduce HCI contributions around novel interactions and technical innovations~\cite{wobbrock2016research}, without becoming exclusively reliant on them. While leveraging the unprecedented capabilities of LLMs in text generation, to overcome limitations of existing technical options in processing symbol motions, we also trained new adapter models on small, task-specific datasets and manipulated the internal workings of pretrained models at a layer beyond just prompting.
Through this approach, \sys{} could even outperform one of the most advanced and large general-purpose vision-language models with custom adapter models and smaller LLM trained on far fewer resources (e.g., GPUs, the dataset). We hope that this work provides a template for how the HCI community could potentially drive interactional and technical innovations in human-AI interaction research, instead of only building around the capabilities of existing models. 
For instance, HCI researchers can focus on intelligent haptic understanding by collecting contextualized datasets and custom-training models.
One realistic barrier would be the lack of GPU resources. For that, one technical option can be considering parameter efficient training approaches~\cite{han2024parameterefficientfinetuning}, which require far fewer resources than training models from scratch. 

While our technical solution made improvements over existing large AI models, we acknowledge that there is still room for improvement. For instance, we used LSTMs in many parts of our system, as they are technically simpler and more efficient than other options (such as transformers). However, as researchers keep improving transformers~\cite{leviathan2023fast}, adopting their best practices might result in additional improvements. Moreover, for embedding motions into action embedding in SBERT space, we used mean absolute error as loss, but contrastive loss~\cite{radford2021learningtransferable} could be another option. We used our approach as we do not have a lot of motion-text data pairs, but if we do have more data, contrastive loss can be an option. For the motion generation, we used mean squared error as loss, but it might not be the most desirable for generative tasks. This loss would assume training data as the only ``gold'' motions and might limit diversity in generations. For future work, we can learn from previous work on similar generative tasks~\cite{ha2018aneural} and use their loss functions to improve motion generation. In motion-conditioned story text generation, our evaluation showed that our approach of interpolating top k actions had minimal benefit over only considering the top 1 action. Improving models with richer datasets (e.g., with soft action labels instead of discrete labels) could result in further benefits of our approach. Moreover, making improvements related to user study participants' suggestions, such as resolving biases and allowing complex scenes, could be technical future work.

%% file: sections/09_conclusion.tex
\section{Conclusion}

We present \sys{}, which facilitates human-AI visual storytelling via toy-playing interactions. As people can anthropomorphize motions of character symbols, toy-playing interaction can be not only a good means to express intent about character-character interactions but also a form of story content in its own right. We enabled toy-playing interaction by training models that translate symbol motion and story text into translational action vectors so that we can generate texts and motions conditioned on those vectors.
With a technical evaluation, we show that \sys{} outperforms the strong baseline, GPT-4o, in action recognition from motions, motion-conditioned story text generation, and action-conditioned motion generation. In a user study, participants used toy-playing interaction for expressing under-developed intentions that are hard to verbalize. However, due to the unique motion-focused design of this modality, toy-playing interactions are abstract in their interpretations. Hence, participants found it ideal to complement it with natural language prompting, which allows expressing concrete intentions. With discussion on the design space, implications, and technical aspects of toy-playing interactions, we hope this work sheds light on enabling novel multimodal interactions for AI-powered storytelling.

%% file: sections/10_appendix.tex
\section{Technical Details}

\subsection{Prompt: \texttt{action+char2text}}
\label{sec:prompt_actionchar2text}
\begin{framed}
    \footnotesize \noindent My story has the following characters: \\
    - \texttt{\{first character's name\}}: \texttt{\{first character's description\}} \\
    - \texttt{\{second character's name\}}: \texttt{\{second character's description\}} \\ \\
    The overall description of the scene is given below: \\
    \texttt{\{scene description\}} \\ \\
    \texttt{\{if previous story sentences exist\}}

    \hangindent=1.5em
    \hangafter= 1
    The previous story look like below: 
    \texttt{\{previous story sentences\}}

    \noindent \texttt{\{end if\}} \\ \\
    \texttt{\{if following story sentences exist\}}

    \hangindent=1.5em
    \hangafter= 1
    Note that after the sentence you are going to write, below sentences will follow. That is, your sentence will come before the following sentences: \texttt{\{following story sentences\}}

    \noindent \texttt{\{end if\}} \\ \\
    $-----$ \\
    Task: \\
    Without preamble, write a story sentence that continues the previous story into an interesting way. The sentence should be fewer than 30 words. In the story sentence to write, \texttt{\{first character's name\}} should actively do the below action, while \texttt{\{second character's name\}} usually being the subject of \texttt{\{first character's name\}}'s action (e.g., for 'throw', \texttt{\{first character's name\}} throws \texttt{\{second character's name\}}, for 'approach', \texttt{\{first character's name\}} approaches \texttt{\{second character's name\}}). \\ 
    \texttt{\{if top $k$ base actions have one of [``avoid,'' ``escape,'' ``ignore,'' ``leave'']\}}
    
        \hangindent=1.5em
        \hangafter= 1
        Only if the action is close to [escape, leave, avoid, ignore], \texttt{\{second character's name\}} should cause \texttt{\{first character's name\}} to take the action, like chasing or bothering \texttt{\{first character's name\}}.

    \noindent \texttt{\{end if\}} \\
    Action: \texttt{\{soft action prompt\}} \\ \\
    \texttt{\{if the user provided a natural language prompt\}}

    \hangindent=1.5em
    \hangafter= 1
    Try to consider following instruction also when writing the sentence (but prioritize the action that characters should take): \texttt{\{the natural language prompt\}}

    \noindent \texttt{\{end if\}}    
\end{framed}

\subsection{Prompt: \texttt{text2char}}
\label{sec:prompt_text2char}
\begin{framed}
    \footnotesize \noindent My story has the following characters:\\
    - character\_0: \texttt{\{first character's name\}}\\
    - character\_1: \texttt{\{second character's name\}}\\
    Consider this sentence: \texttt{\{story sentence\}}\\ \\
    Regarding this sentence, which character is taking the following action?\\
    Action: \texttt{\{soft action prompt\}}\\ \\
    Answer with a number 0 or 1, without any preamble.
\end{framed}

\section{Evaluation Prompts}

\subsection{\texttt{GPT-4o-V} Prompt for vs. \texttt{motion2action}}
\label{sec:gpt4ov-prompt-motion2action}

\begin{framed}
    \footnotesize \noindent The following sequence of image is describing an action described in character symbols denoted in white and black triangles. The frame rate of image sequence is 10 frames per second. Rank the following actions in the order of their relevance to the sequence of images: \\
    encircle, poke, examine, pull, push, escape, leave, kiss, fight, throw, hug, argue with, flirt with, play with, avoid, lead, herd, mimic, scratch, capture, chase, block, ignore, bother, accompany, approach, follow, huddle with, creep up on, hit, talk to, tickle \\ \\
    Try to rank all the actions, with relative weights assigned to each action. Answer without any preamble or any reasoning, and in the following format (as an example): \\ \\
    1. encircle - 0.5 \\
    2. poke - 0.3 \\
    3. examine - 0.2 \\
    ... \\ \\
    \texttt{\{Series of images appended\}}
\end{framed}

\subsection{\texttt{GPT-4o-C} Prompt for vs. \texttt{motion2action}}
\label{sec:gpt4oc-prompt-motion2action}
\begin{framed}
    \footnotesize \noindent The following sequence of coordinates is describing an action described in character symbols. \\ \\
    \texttt{\{sequence of coordinates in ((x1, y1, r1), (x2, y2, r2))\}}\\ \\ 
    For each frame, the first item is for the first character symbol, and the second item is for the second character symbol. For each character symbol, the first element is the x position, the second is the y position, and the third is rotation in radian. The frame rate of image sequence is 10 frames per second. Rank the following actions in the order of their relevance to the sequence of images: \\
    encircle, poke, examine, pull, push, escape, leave, kiss, fight, throw, hug, argue with, flirt with, play with, avoid, lead, herd, mimic, scratch, capture, chase, block, ignore, bother, accompany, approach, follow, huddle with, creep up on, hit, talk to, tickle \\ \\
    Try to rank all the actions, with relative weights assigned to each action. Answer without any preamble or any reasoning, and in the following format (as an example): \\ \\
    1. encircle - 0.5 \\
    2. poke - 0.3 \\
    3. examine - 0.2 \\
    ...
\end{framed}

\subsection{\texttt{GPT-4o-V} Prompt for vs. \texttt{motion2char}}
\label{sec:gpt4ov-prompt-motion2char}
\begin{framed}
    \footnotesize \noindent The following sequence of image is describing an action described in character symbols denoted in white and black triangles. The described action is \texttt{\{action\}}. The frame rate of image sequence is 10 frames per second. Decide which character is 'actively' taking the action. That is, who is more 'active' actor that influence the counterpart as the subject of the action? If difficult to decide, answer with a bit more plausible one. Answer with 1 if the black triangle is the active one and with 0 if the white triangle is the active one. Answer without any preamble or any reasoning. \\ \\
    \texttt{\{A series of images appended that comprises the motion in 10 frames per second\}}
\end{framed}

\subsection{\texttt{GPT-4o-C} Prompt for vs. \texttt{motion2char}}
\label{sec:gpt4oc-prompt-motion2char}
\begin{framed}
    \footnotesize \noindent The following sequence of coordinates is describing an action described in character symbols. \\ \\
    \texttt{\{A sequence of coordinates in ((x1, y1, r1), (x2, y2, r2)) that comprises the motion in 10 frames per second\}} \\ \\
    For each frame, the first item is for the first character symbol, and the second item is for the second character symbol. For each character symbol, the first element is the x position, the second is the y position, and the third is rotation in radian. The described action is \texttt{\{action\}}. The frame rate of image sequence is 10 frames per second. Decide which character is 'actively' taking the action. That is, who is more 'active' actor that influence the counterpart as the subject of the action? If difficult to decide, answer with a bit more plausible one. Answer with 0 if the first character is the active one, and with 1 if the second one is the active one. Answer without any preamble or any reasoning.
\end{framed}

\subsection{Prompt to Generate Story Settings}
\label{sec:generating-story-settings}
\begin{framed}
    \footnotesize \noindent Create 10 sets of settings where each setting includes two characters' names and descriptions. Description should be at most two sentences. For each setting, use the following format.\\ \\ 
    -Setting: [Setting] \\
    -[Name 1]: [Description 1] \\
    -[Name 2]: [Description 2] \\ \\
    Answer without any preamble.
\end{framed}

Settings generated with GPT-4o are like below:
\begin{framed}
    \footnotesize \noindent \textbf{Setting 1} \\
    Scene: Enchanted Forest \\
    Character 1: Ariana - A skilled archer with a mysterious past, Ariana is known for her sharp eyes and even sharper wit.\\
    Character 2: Thorn - A brooding warrior with a heart of gold, Thorn protects the forest with his life.\\ \\
    \textbf{Setting 2}\\
    Scene: Space Station Omega \\
    Character 1: Captain Zara - A fearless leader with a knack for strategy, Zara commands respect from her crew. \\
    Character 2: Dr. Lin - A brilliant scientist who specializes in alien biology, Lin is always on the brink of a groundbreaking discovery. \\ \\
    \textbf{Setting 3} \\
    Scene: Medieval Kingdom \\
    Character 1: Sir Cedric - A noble knight with a strong sense of justice, Cedric is the king's most trusted advisor. \\
    Character 2: Lady Elara - A cunning diplomat, Elara uses her charm and intelligence to navigate the treacherous waters of court politics. \\ \\
    \textbf{Setting 4} \\
    Scene: Post-Apocalyptic City \\
    Character 1: Raven - A resourceful scavenger with a dark sense of humor, Raven knows every nook and cranny of the ruined city. \\
    Character 2: Jax - A former soldier turned mercenary, Jax is haunted by the ghosts of his past. \\ \\
    \textbf{Setting 5} \\
    Scene: Underwater Research Facility \\
    Character 1: Dr. Marina - An oceanographer with a passion for marine life, Marina is dedicated to uncovering the secrets of the deep. \\
    Character 2: Captain Finn - A seasoned diver and the facility's security chief, Finn is always ready to face the unknown. \\ \\
    \textbf{Setting 6} \\
    Scene: Victorian London \\
    Character 1: Inspector Grey - A sharp-minded detective with a keen eye for detail, Grey is relentless in his pursuit of justice. \\ 
    Character 2: Miss Lila - A talented inventor and Grey's confidante, Lila uses her gadgets to aid in solving crimes. \\ \\
    \textbf{Setting 7} \\
    Scene: Cyberpunk Metropolis \\
    Character 1: Neo - A hacker with a rebellious streak, Neo fights against the oppressive corporations that control the city. \\
    Character 2: Vex - A street-smart informant, Vex has connections in every corner of the metropolis. \\ \\
    \textbf{Setting 8} \\
    Scene: Ancient Egypt \\
    Character 1: High Priestess Nefertari - A powerful and wise spiritual leader, Nefertari is revered by her people.\\
    Character 2: General Horus - A loyal and brave warrior, Horus is the Pharaoh's right hand in both battle and strategy. \\ \\
    \textbf{Setting 9} \\
    Scene: Fantasy Academy \\
    Character 1: Professor Eldric - A master of elemental magic, Eldric is both feared and respected by his students. \\
    Character 2: Luna - A prodigious young witch with a mysterious lineage, Luna is determined to prove herself. \\ \\
    \textbf{Setting 10} \\
    Scene: Haunted Mansion \\
    Character 1: Madame Seraphine - A medium with a tragic past, Seraphine communicates with the spirits that haunt the mansion. \\ 
    Character 2: Jasper - A skeptical journalist, Jasper is determined to uncover the truth behind the mansion's eerie occurrences.
\end{framed}

\subsection{\texttt{GPT-4o-V} Prompt for Text Generation.}
\label{sec:gpt4ov-prompt-textgen}
\begin{framed}
    \footnotesize \noindent The following sequence of images is describing an action described in character symbols denoted in white and black triangles. The frame rate of image sequence is 10 frames per second. Write down the story sentence that would be well expressed with the sequence of the images. \\ \\
    Black triangle is a character named \texttt{\{first character's name\}}, with the following description: \texttt{\{first character's description\}}. \\ \\
    White triangle is a character named \texttt{\{second character's name\}}, with the following description: \texttt{\{second character's description\}}. 
    The scene of the story is \texttt{\{scene/setting sentence\}}. Write the story sentence without preamble. \\ \\
    \texttt{\{A series of images that comprises the motion in 10 frames per second\}}
\end{framed}

\subsection{\texttt{GPT-4o-C} Prompt for Text Generation.}
\label{sec:gpt4oc-prompt-textgen}
\begin{framed}
    \footnotesize \noindent The following sequence of coordinates is describing an action described in character symbols. \\ \\
    \texttt{\{A sequence of coordinates in ((x1, y1, r1), (x2, y2, r2)) that comprises the motion in 10 frames per second\}} \\ \\ 
    For each frame, the first item is for the first character symbol, and the second item is for the second character symbol. For each character symbol, the first element is the x position, the second is the y position, and the third is rotation in radian. The frame rate of image sequence is 10 frames per second. Write down a story sentence that would be well expressed with the sequence of the images. \\ \\
    The first character is named \texttt{\{first character's name\}}, with the following description: \texttt{\{first character's description\}}. \\ \\
    The second character is named \texttt{\{second character's description\}}, with the following description: \texttt{\{second character's description\}}. The scene of the story is \texttt{\{scene/setting sentence\}}. Write the story sentence without preamble.
\end{framed}

\subsection{\texttt{GPT-4o-Pro} Prompt for vs. \texttt{proactive action+char2motion}}
\label{sec:gpt4pro-prompt-motiongen}

\begin{framed}
    \footnotesize \noindent The following sequence of coordinates is describing an action described in character symbols.\\ \\ 
    \texttt{\{A sequence of previous coordinates in ((x1, y1, r1), (x2, y2, r2)) that comprises the motion in 10 frames per second\}} \\ \\
    For each frame, the first item is for the first character symbol, and the second item is for the second character symbol. For each character symbol, the first element is the x position, the second is the y position, and the third is rotation in radian. The frame rate of image sequence is 10 frames per second. Through this animation, the \texttt{\{first/second\}} character should perform the following action to the \texttt{\{second/first\}} character: \texttt{\{action\}}. Without preamble, generate the first character's possible next x position, y position, and rotation. The format should look like (x, y, r) and the generated frame should denote the targeted action.  Only give the first character's next position and rotation, without any other reasoning or rationales.
\end{framed}

\subsection{\texttt{GPT-4o-Re} Prompt for vs. \texttt{reactive action+char2motion}}
\label{sec:gpt4re-prompt-motiongen}

\begin{framed}
    \footnotesize \noindent The following sequence of coordinates is describing an action described in character symbols. \\ \\ 
    \texttt{\{A sequence of previous coordinates in ((x1, y1, r1), (x2, y2, r2)) that comprises the motion in 10 frames per second\}} \\ \\
    For each frame, the first item is for the first character symbol, and the second item is for the second character symbol. For each character symbol, the first element is the x position, the second is the y position, and the third is rotation in radian. The frame rate of image sequence is 10 frames per second. Through this animation, the \texttt{\{first/second\}} character should perform the following action to the \texttt{\{second/first\}} character: \texttt{\{action\}}. For the upcoming frame, the second character's coordinate should look like following: (\texttt{\{The upcoming coordinate of the character whose motion is not being generated, in (x, y, r)\}}). Without preamble, generate the first character's possible next x position, y position, and rotation. The format should look like (x, y, r) and the generated frame should denote the targeted action. Only give the first character's next position and rotation, without any other reasoning or rationales.
\end{framed}

\section{Technical Evaluation Interface}
\label{sec:annotation_task}

\begin{figure}[h]
  \includegraphics[width=0.478\textwidth]{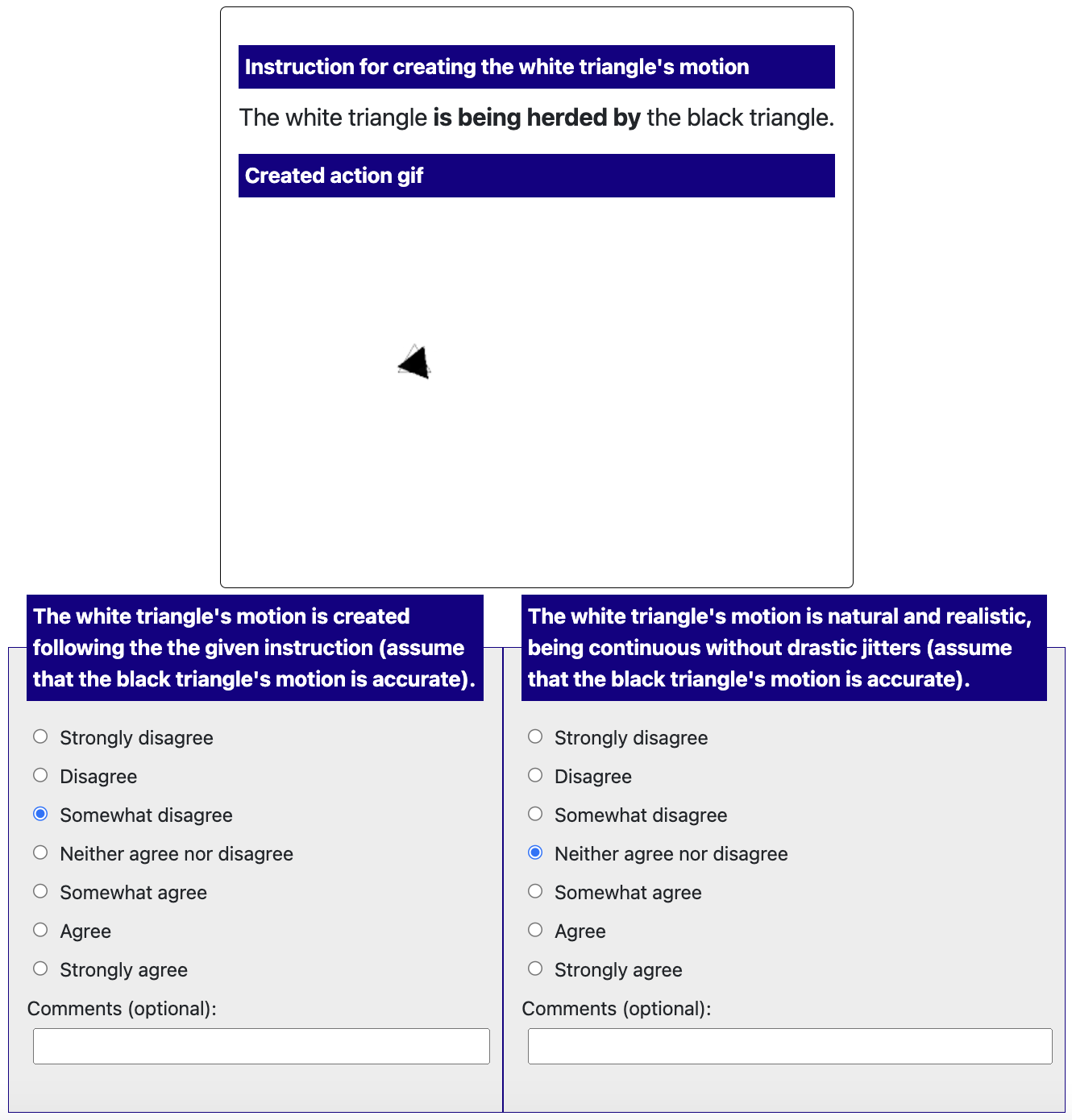}
  \caption{The interface to evaluate the quality of action-conditioned motion generation.}
  \label{fig:motion_annotation_task}
\end{figure}

\begin{figure*}[h]
  \includegraphics[width=\textwidth]{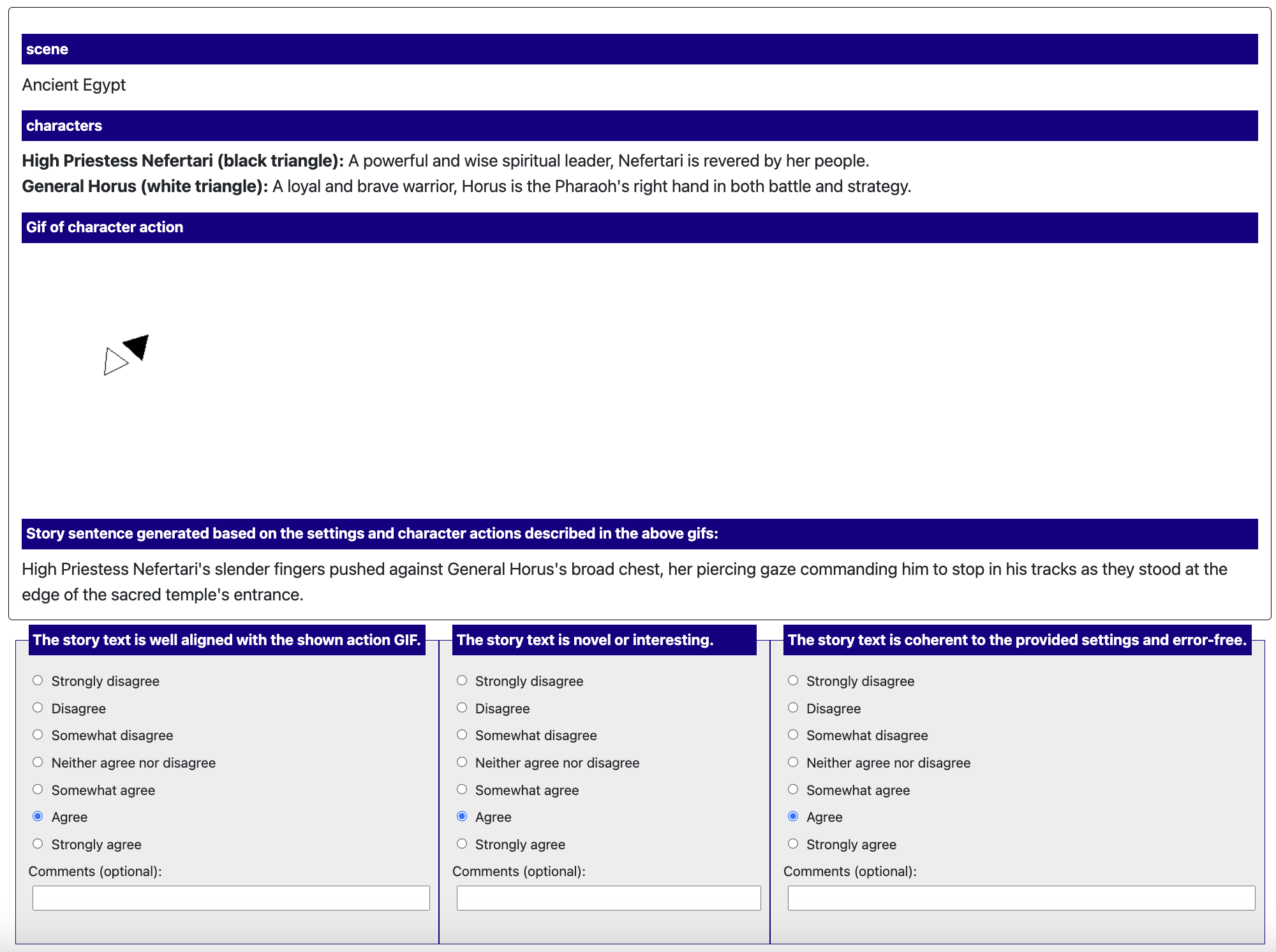}
  \caption{The interface to evaluate the quality of motion-conditioned text generation.}
  \label{fig:text_annotation_task}
\end{figure*}

We provide annotation task interfaces for motion-conditioned text generation and action-conditioned motion generation in Figure~\ref{fig:text_annotation_task} and~\ref{fig:motion_annotation_task}, respectively.